\documentclass[aps,prl,twocolumn,superscriptaddress]{revtex4-1}

\usepackage{graphicx} 
\usepackage{dcolumn} 
\usepackage{amsmath, amssymb, amsbsy}
\usepackage{latexsym}
\usepackage{epsfig}
\usepackage{subfigure}
\usepackage{epstopdf}
\usepackage{makecell}
\usepackage{color}
\usepackage{array}
\usepackage{txfonts}
\usepackage{bm, bbm}
\usepackage{graphicx}
\usepackage{appendix}
\usepackage{diagbox, multirow}
\usepackage[percent]{overpic}
\usepackage[colorlinks=true,linkcolor=blue,citecolor=blue,urlcolor=blue,bookmarks=false]{hyperref}

\renewcommand{\d}{\mathrm{d}}
\newcommand{\pt}{\partial}
\newcommand{\e}{\mathrm{e}}
\newcommand{\m}[1]{\mathrm{#1}}
\renewcommand{\bf}[1]{\mathbf{#1}}
\renewcommand{\cal}[1]{\mathcal{#1}}

\newcommand{\pp}[2]{\frac{\partial{#1}}{\partial{#2}}}

\renewcommand{\Im}{\mathrm{Im}}

\newcommand{\ex}{\hat{\bf{x}}}
\newcommand{\ey}{\hat{\bf{y}}}
\newcommand{\ez}{\hat{\bf{z}}}
\newcommand{\sn}{\,\mathrm{sn}}
\newcommand{\cn}{\,\mathrm{cn}}
\newcommand{\dn}{\,\mathrm{dn}}

\renewcommand{\a}{\alpha}
\renewcommand{\b}{\beta}
\newcommand{\g}{\gamma}

\newcommand{\de}{\delta}
\newcommand{\De}{\Delta}
\newcommand{\eps}{\epsilon}

\newcommand{\z}{\zeta}

\renewcommand{\l}{\lambda}
\newcommand{\s}{\sigma}

\newcommand{\vp}{\varphi}
\newcommand{\om}{\omega}

\newcommand{\br}{\mathbf{r}}
\newcommand{\bR}{\mathbf{R}}

\newcommand{\Hpara}{H_{\parallel}}
\newcommand{\Hperp}{H_{\perp}}

\newcommand{\wpara}{\om_{\m{c}\parallel}}
\newcommand{\wperp}{\om_{\m{c}\perp}}
\newcommand{\lpara}{\ell_{\parallel}}
\newcommand{\lperp}{\ell_{\perp}}
\newcommand{\ha}{\hat{a}}
\newcommand{\had}{\hat{a}^\dag}
\newcommand{\hb}{\hat{b}}
\newcommand{\hbd}{\hat{b}^\dag}
\newcommand{\tpsi}{\tilde{\psi}}
\newcommand{\fed}{\mathcal{F}}

\begin{document}
\bibliographystyle{apsrev4-1}

\title{Inhomogeneous superconducting states in two weakly linked superconducting ultrathin films}

\author{Gao-Wei Qiu}
\affiliation{Kavli Institute for Theoretical Sciences, University of Chinese Academy of Sciences, Beijing 100190, China}

\author{Yi Zhou}
\email{yizhou@iphy.ac.cn}
\affiliation {Beijing National Laboratory for Condensed Matter Physics $\&$ Institute of Physics, Chinese Academy of Sciences, Beijing 100190, China}
\affiliation{Songshan Lake Materials Laboratory, Dongguan, Guangdong 523808, China}
\affiliation{Kavli Institute for Theoretical Sciences, University of Chinese Academy of Sciences, Beijing 100190, China}
\affiliation{CAS Center for Excellence in Topological Quantum Computation, University of Chinese Academy of Sciences, Beijing 100190, China}
\date{\today}

\begin{abstract}
	A sufficiently large parallel magnetic field will generate staggered supercurrent loops and superfluid density wave in two weakly linked superconducting (SC) ultrathin films, resulting in an inhomogeneous Fulde-Ferrell-Larkin-Ovchinnikov (FFLO) state. The SC order parameter of such an FFLO state is characterized by Bloch wave functions, called the ``Bloch SC state''. The staggered supercurrent loops form an array of Josephson vortex-antivortex pairs, instead of the usual Josephson vortex lattice. Enclosing a unit cell of the array, the London's fluxoid is quantized as $\Phi^{\prime}=\Phi_0=hc/2e$, while the net orbital magnetization caused by the staggered supercurrent is zero. Meanwhile, a small parallel magnetic field gives rise to an Fulde-Ferrell (FF) state that has uniform superfluid density. The phase transition between the Bloch SC state and the FF state belongs to the universality class of two-dimensional commensurate-incommensurate transitions. An analytical solution in terms of Jacobian elliptic functions is found to be an excellent approximation to the Bloch SC order parameter.
\end{abstract}
\maketitle

Inhomogeneous superconductivity that breaks translational symmetry spontaneously has been attracting growing attention from diverse fields in physics, ranging from condensed matter to high-energy physics~\cite{RMP04}. Quintessential examples for such inhomogeneous superconductors include the well-known Fulde-Ferrell-Larkin-Ovchinnikov (FFLO) state~\cite{FFstate,LOstate} and its generalized version, the pair density wave (PDW) state~\cite{ARCMP20}. The FFLO or PDW state is a superconducting state with a nonuniform superconducting (SC) order parameter. Such an inhomogeneous SC state was proposed as a mother state of other ordering states. For instance, the partial melting of the PDW can give rise to a charge density wave (CDW) order, a uniform charge-$4e$ SC order, and a loop current order. Moreover, the PDW is expected to host fantastic quasiparticle excitations as well as topological defects, such as in-gap Bogoliubov quasiparticles in an $s$-wave superconductor and a half-flux ($hc/4e$) vortex bound with a CDW dislocation~\cite{ARCMP20}.

Meanwhile, extensive research activities in condensed matter physics and material sciences have been devoted to superconducting thin films and layered superconductors over the past several decades~\cite{Eom,Morshedloo,Bulaevski}. Indeed, layered superconductors can be viewed as intrinsic superconductors with weak links, namely, adjacent superconducting layers couple each other via Josephson junctions~\cite{Josephson1962,Josephson1965,RMP79,Kleiner1992,Kleiner1994}. Among these research objects, two-dimensional (2D) SC systems in the presence of an in-plane magnetic field \cite{Agterberg,Barzykin,Aoyama} is of particular interest, on which the emergence of unconventional superconductivity due to the applied magnetic field and spin-orbit coupling effect was proposed~\cite{Dimitrova,Aoyama,Barzykin}. It has been suggested that an inhomogeneous FFLO state can be induced by an in-plane magnetic field in bilayer transition metal dichalcogenides (TMDs)~\cite{bilayerTMD}, such as $\m{MoS}_2$~\cite{Lu,Yu} and $\m{NbSe}_2$~\cite{Xi}. Very recently, with the help of a quasiparticle interference (QPI) technique, the experimental observation of a segmented Fermi surface inside the superconducting energy gap was reported in $\m{Bi}_2\m{Te}_3$ thin films proximitized by the superconductor $\m{NbSe}_2$ and under an in-plane magnetic field, which indicates the existence of a PDW or FFLO state~\cite{Zheng21}.

In this Letter, we study a model for two weakly coupled SC ultrathin films in an applied parallel magnetic field, which allows us to explore various PDW states. The model is gauge invariant and essentially equivalent to the Lawrence-Doniach model~\cite{LawrenceDoniach,YangKun} in the double-layer limit. 

\begin{figure}[tb]
	\centering
	\vspace{5pt}
	\subfigure{
		\label{model}
		\begin{overpic}[width=\linewidth]{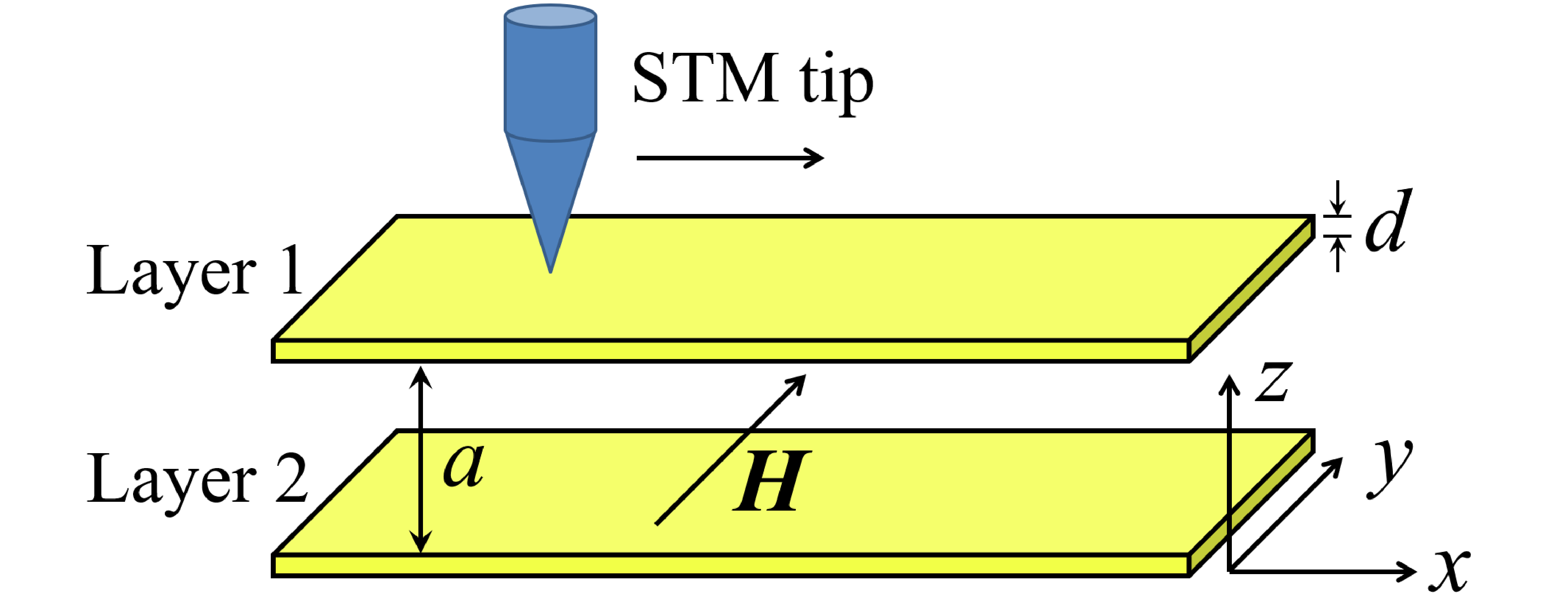}
			\put(0,36){(a)}
		\end{overpic}
	}
	\vfill
	\subfigure{
		\label{density18}
		\begin{overpic}[width=\linewidth]{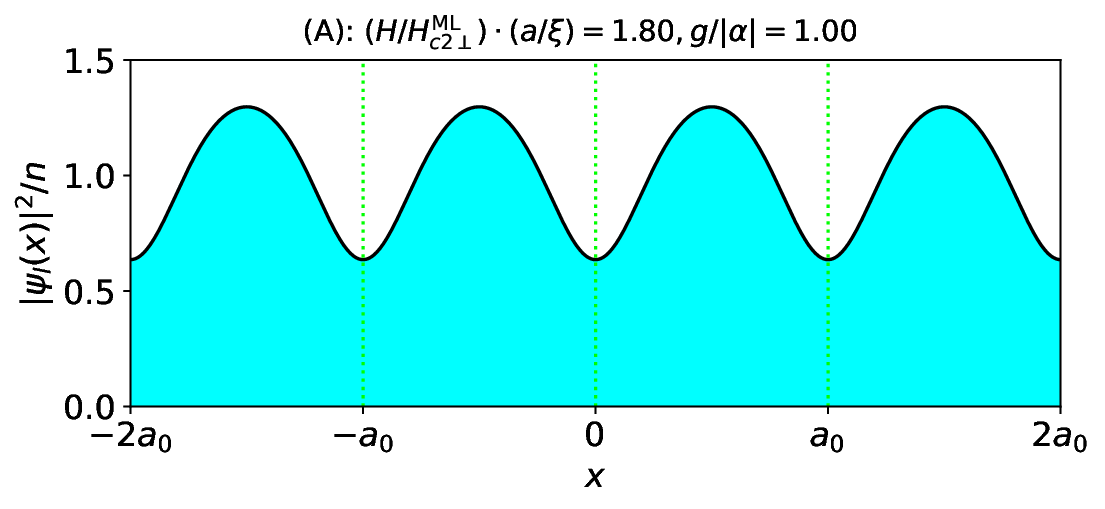}
			\put(0,42){(b)}
		\end{overpic}
	}
	\vfill
	\subfigure{
		\label{density40}
		\begin{overpic}[width=\linewidth]{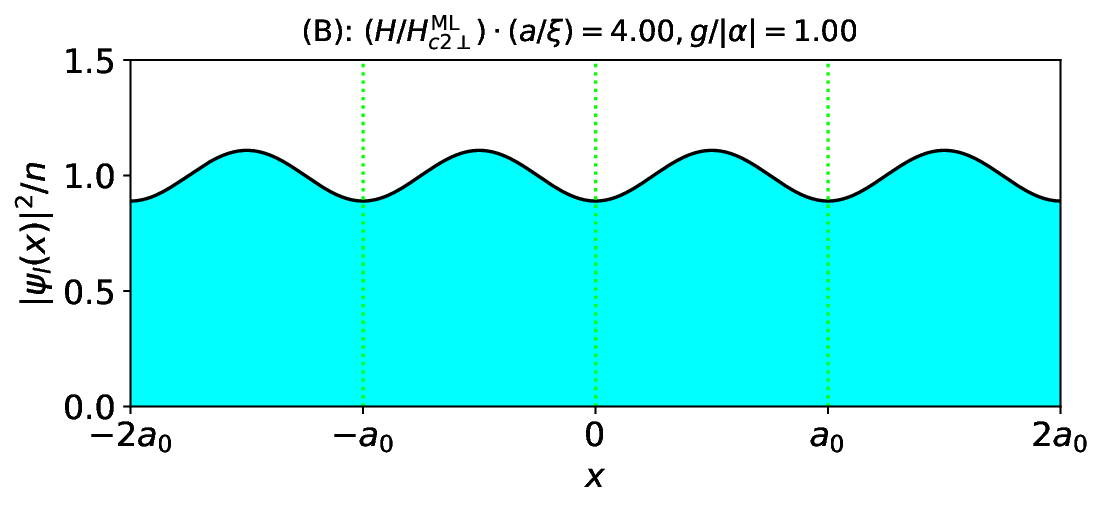}
			\put(0,42){(c)}
		\end{overpic}
	}
	\caption{(a) A bilayer SC system in an applied parallel magnetic field. The superfluid density wave can be probed by STM. Spatial modulation of the superfluid density for states (b) A and (c) B in the phase diagram (see Fig.~\ref{phasediag})}
	\label{fig1}
\end{figure}

\emph{Model.}
We consider two layers of superconducting ultrathin films in the presence of an applied parallel magnetic field. These two layers are weakly linked to each other such that the Josephson tunneling current can flow in the direction perpendicular to them. The thickness of each layer $d$ is considerably small compared to the penetration depth $\lambda$ and coherence length $\xi$ of the superconductor, so that a diamagnetic current loop can not be induced by the applied in-plane magnetic field within each layer. 

Assuming the separation between the two layers is $a$, we set up the coordinate system as depicted in Fig.~\ref{model}, on which the external magnetic field is along the $y$ direction, i.e., $\bf{H}=H\ey$. The corresponding vector potential is given by $\bf{A}=Hz\ex$ in the Landau gauge choice. 
Thus, the system is described by a  two-component gauge-invariant Ginzburg-Landau (GL) free-energy functional~\cite{Ginzburg-Landau} as follows,
\begin{align}
	f[\psi_1(\br),&\,\psi_2(\br)] = f_n + \sum_{l=1,2}\bigg\{\a|\psi_l(\br)|^2 + \frac{\b}{2}|\psi_l(\br)|^4 \notag\\
	&+ \frac{1}{2m^*}\left|\left(\frac{\hbar}{i}\bm{\nabla} - \frac{e^*}{c}\bf{A}_l(\br)\right)\psi_l(\br)\right|^2\bigg\} \notag\\ 
	&+ g\left[\psi_1(\br)^*\psi_2(\br)\,\e^{i\frac{2\pi}{\Phi_0}\int_{2}^{1}\bf{A}\cdot d\bf{s}} +\m{c.c.}\right]. \label{eq:f1}
\end{align}
Here, $l=1,2$ labels two layers, $\psi_1(\bf{r})$ and $\psi_2(\bf{r})$ are superconducting order parameters on them, and $\bf{r}=(x,y)$ is the 2D Cartesian coordinate. Since the thickness of each thin film $d\ll\lambda,\xi$, it is reasonable to presume that the order parameters are independent of $z$. Defining $\eta_1=+1$ and $\eta_2=-1$, the vector potential on each layer reads $\bf{A}_l(\br)=\eta_l(Ha/2)\ex$.  The last  term in Eq.~\eqref{eq:f1} describes the Josephson tunneling effect, where $g$ is the Josephson coupling energy. The orbital effect of the external magnetic field is introduced by the Peierls phase factor $\e^{i\frac{2\pi}{\Phi_0}\int_{2}^{1}\bf{A}\cdot d\bf{s}}$, which ensures the gauge invariance of the GL free energy. $\Phi_0=hc/e^*=hc/2e$ is the flux quantum. 

Note that our model is applicable to a superconductor-insulator-superconductor (S-I-S) junction, as long as the thickness of each superconducting layer, $d\ll \lambda,\xi$. Such an S-I-S junction can be realized by the growth of a superconducting layer on both surfaces of an insulating thin film to form a sandwich structure. Other possible material realizations include but are not limited to superconducting TMDs.

The 2D model given in Eq.~\eqref{eq:f1} can be further simplified.  First, we consider the local tunneling between the two layers only, so that $\int_{2}^{1}\bf{A}\cdot d\bf{s}=0$ in the Landau gauge. Second, the Landau gauge choice gives rise to an explicitly $y$-independent form of GL free energy. Moreover, possible minima of the GL free-energy functional are always given by $\partial_{y}\psi_{l=1,2}=0$. Thus, the original 2D model can be simplified by the substitution $\psi_{l}(\br)\to\psi_{l}(x),$ and the GL free-energy functional will be reduced to the one-dimensional form as follows, 
\begin{align}\label{freeEne}
	f[\psi_1,\psi_2] = f_n &+ \sum_{l=1,2}\bigg\{\a|\psi_l|^2 + \frac{\b}{2}|\psi_l|^4 \notag\\
	&+ \frac{\hbar^2}{2m^*}\left|\left(\pp{}{x} - i\eta_l\frac{k_0}{2}\right)\psi_l\right|^2\bigg\} \notag\\
	&+ g\left(\psi_1^*\psi_2+\psi_2^*\psi_1\right),
\end{align}
where a magnetic field-dependent wave vector $k_0=2\pi Ha/\Phi_0$ is introduced, and the corresponding length scale reads $a_0={}2\pi/k_0=\Phi_0/Ha$. The convenience of these notations will be seen later. Taking variations with respect to $\psi_l^*$ in Eq.~\eqref{freeEne}, we obtain coupled GL equations,
\begin{equation}\label{GLeq}
	\a\psi_l + \b|\psi_l|^2\psi_l - \frac{\hbar^2}{2m^*}\left(\pp{}{x}-i\eta_l\frac{k_0}{2}\right)^2\psi_l + g\psi_{\bar{l}} = 0,
\end{equation}
where $l,\bar{l}=1,2$, and $\bar{l}$ represents the opposite layer to $l$. 

\emph{Symmetry.} First of all, let us discuss relevant symmetry operations acting on coupled GL equations. The time reversal $\cal{T}$ acts as: $H\to{}-H,\psi_{l}(x)\to\psi_{l}(x)^{*}$, the spatial reflection $\cal{P}_{z}$ acts as $l\to\bar{l}$, so that the joint operation $\cal{T}\cal{P}_{z}$ acts as $\psi_{l}(x)\to\psi_{\bar{l}}(x)^{*}$. It is remarkable that Eqs.~\eqref{GLeq} keep invariant under the operation $\cal{T}\cal{P}_{z}$ followed by the complex conjugate, thereby allowing a $\cal{T}\cal{P}_{z}$ symmetric solution, $\psi_{\bar{l}}(x)^{*}=\psi_{l}(x)$.

We begin with some exact solutions to Eqs.~\eqref{GLeq}. It is easy to see that the coupled GL equations \eqref{GLeq} have a trivial but exact solution, which is a Fulde-Ferrell (FF) state~\cite{FFstate} indeed. Meanwhile, Eqs.~\eqref{GLeq} will become decoupled at $g=0$, and give rise to another exact solution, called the decoupled SC state. Below we shall examine these two solutions:

(i) FF state. This state is given by a pair of constant order parameters on two SC layers, which takes the form
\begin{equation}\label{FFstate}
	\psi_{l}(x) =  \sqrt{\rho_\m{s}^\m{FF}}\,\e^{i\vp_l},
\end{equation}
where $\rho_\m{s}^\m{FF}=-\b^{-1}\left(\a+\eps_H-|g|\right)$, and $\eps_H=\hbar^2k_0^2/8m^*$ is an energy associated with the magnetic field $H$. Such a ground state carries diamagnetic supercurrent flow, and by definition, it is an FF state~\footnote{At first sight, the SC order parameter $\psi_{l}(x) $ does not vary spatially and suggests a uniform state rather than an FF state, which is true in the absence of a magnetic field. However, the situation changes in the presence of a magnetic field, because the SC order parameter does depend on the gauge choice and we should consider the gauge invariant physical observable, say, the current, instead. Indeed, this state carries finite and opposite supercurrents in both layers, so that it is an FF state by definition.}. The phase difference $\De\vp\equiv\vp_1-\vp_2$ is determined as follows: (1) $\Delta\varphi=\pi$ for $g>0$ and (2) $\De\vp=0$ for $g<0$. The free-energy density reads $\fed_\m{FF}=-\b^{-1}\left(\a+\eps_H-|g|\right)^2$.
The non-negative constraint to the superfluid density $\rho_\m{s}^\m{FF}$ requires an upper bound for the external magnetic field $H$,
\begin{equation}\label{Hcstar}
	H_\m{c}^* = \frac{\Phi_0}{\pi a\xi} \left(1+\frac{2m^*\xi}{\hbar^2}|g|\right)^{1/2},
\end{equation}
where the superconductor coherence length $\xi$ is determined by the relation $|\a|=\hbar^2(2m^*\xi^2)^{-1}$. A nonzero FF state solution is not allowed when $H$ exceeds $H_\m{c}^* $.

(ii) Decoupled SC state. When $g=0$, Eqs.~\eqref{GLeq} have a plane-wave solution as follows, 
\begin{equation}\label{DCstate}
	\psi_l(x) = \sqrt{\rho_{\m{s}0}}\,\e^{i\eta_l\frac{k_0}{2}x},
\end{equation}
where $\rho_{\m{s}0} = -\a/\b$. The free-energy density for this decoupled SC state is $\fed_\m{D}=-\a^2/\b$. The inequality $\fed_\m{D}<\fed_\m{FF}$ always holds as long as  $g=0$ and $H<H_\m{c}^*$. Therefore, the decoupled SC state is energetically favored over the FF state in the absence of the Josephson tunneling. 

In the presence of the Josephson tunneling, say, $g\neq{}0$, the situation will be complicated and more interesting. In this case, the nontrivial exact solution to the nonlinear GL equations \eqref{GLeq} is not available in general. However, we can raise the following issue: Does a SC state exist that is more energetically favored than the decoupled SC state and the FF state? To address this issue, we shall examine the well-known Bloch wave function that is a natural generalization of the plane wave and has been exploited to solve the nonlinear Schr\"{o}dinger equation in the context of cold atoms~\cite{Pethick}. 

\emph{Bloch SC state.}
In order to pursue a lower-energy SC state, we consider a more generic solution which takes the Bloch form as follows,
\begin{equation}\label{ansatz}
	\psi_l(x) = \e^{i\eta_l\frac{k_0}{2}x}\tpsi_l(x),
\end{equation}
where $\tpsi_l(x+a_0)=\tpsi_l(x)$ is a periodic function and $a_0=2\pi/k_0$ as defined before. Following Ref.~\cite{Pethick}, we shall minimize the GL free-energy functional \eqref{freeEne} by expanding $\tpsi_l$ in terms of plane waves,
\begin{equation}\label{expand}
	\tilde{\psi}_l(x) = \sqrt{n}\sum_{\nu=-\infty}^{\infty}a_{l\nu}\e^{i\nu k_0x}.
\end{equation}
where $\nu$ is an integer, and $n=a_0^{-1}\int_{0}^{a_0}dx|\tilde{\psi}_l(x)|^2$ is the average superfluid density on each layer. The coefficients $a_{l\nu}$ are subject to the normalization relation $\sum_{\nu=-\infty}^{\infty}|a_{l\nu}|^2 = 1$. Such a Bloch SC state will become the decoupled SC state when $a_{l\nu}=\delta_{\nu{}0}$.
Evaluating the GL free energy and minimizing it with respect to $a_{l\nu}$ and $n$ result in self-consistent equations as follows~\cite{appendix},
\begin{subequations}\label{discreteGL}
	\begin{align}
		0 &= \left(\frac{\hbar^2k_0^2}{2m^*}\nu^2 +\a\right)a_{l\nu} + ga_{\bar{l},\nu+\eta_l} + n\b\sum_{\nu_1,\nu_2}a_{l\nu_1}^*a_{l\nu_2}a_{l,\nu+\nu_1-\nu_2}, \label{gla} \\
		n &= -\frac{\sum_{l=1,2}\left[\sum_{\nu}\left(\frac{\hbar^2k_0^2}{2m^*}\nu^2 +\a\right)|a_{l\nu}|^2 +  g\sum_{\nu} a_{l\nu}^*a_{\bar{l},\nu+\eta_{l}}\right]}{\b\sum_{l=1,2}\sum_{\nu_1,\nu_2,\nu_3} a_{l\nu_1}^*a_{l\nu_2}^*a_{l\nu_3}a_{l,\nu_1+\nu_2-\nu_3}}. \label{glb}
	\end{align}
\end{subequations}
Note that Eqs.~\eqref{discreteGL} can be derived from Eqs.~\eqref{GLeq} as well.

Before proceeding, we would like to make general remarks on the solutions to nonlinear equations \eqref{discreteGL}: (1) It is more convenient to solve $\{n\beta,a_{1\nu},a_{2\nu}\}$ instead of $\{n,a_{1\nu},a_{2\nu}\}$, such that the parameter $\beta$ will be irrelevant to the solutions.
(2) Without loss of generality, we can set $|\alpha|$ as the energy unit, then the solution $\{n\beta,a_{1\nu},a_{2\nu}\}$ will be determined by two independent parameters $g/|\alpha|$ and $H$. 
(3) For any solution, apart from an overall phase factor, the phase of $a_{l\nu}=|a_{l\nu}|e^{i\phi_{l\nu}}$ can be taken to be $0$ or $\pi$~\footnote{Because $\phi_{l\nu}$'s occur in a free-energy functional in the form of $\cos(\phi_{l_1\nu_1}-\phi_{l_2\nu_2})$ and $\cos(\phi_{l_1\nu_1}-\phi_{l_2\nu_2}+\phi_{l_3\nu_3}-\phi_{l_4\nu_4})$, and a saddle point will be achieved at $\cos(\phi_{l_1\nu_1}-\phi_{l_2\nu_2})=\pm{}1$ and  $\cos(\phi_{l_1\nu_1}-\phi_{l_2\nu_2}+\phi_{l_3\nu_3}-\phi_{l_4\nu_4})=\pm{1}$~\cite{Pethick}, so that we can always choose $a_{l\nu}$ as real numbers.}.
(4) To solve Eqs.~\eqref{discreteGL} numerically, the truncation of the series $\{a_{1\nu}\}$ and $\{a_{2\nu}\}$ has to been introduced, namely, $a_{1\nu}=a_{2\nu}=0$ for $|\nu|>\nu_{\max}$, where $\nu_{\max}$ is a positive integer.
(5) The aforementioned $\cal{T}\cal{P}_{z}$ symmetry, $\psi_{\bar{l}}(x)^{*}=\psi_{l}(x)$, is respected by all the numerically found solutions.

\begin{figure}[tb]
	\centering
	\includegraphics[width=\linewidth]{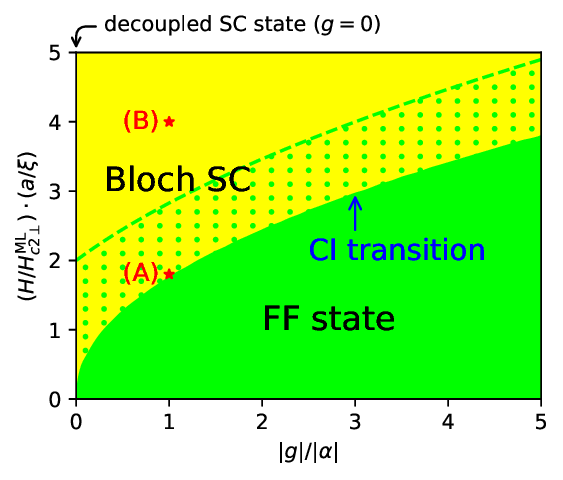}
	\caption{Phase diagram. The Bloch SC state restores the decoupled SC state along the line of $g=0$. The green dashed line in Fig.~\ref{phasediag} indicates the upper bound $H^*_\m{c}$ given in Eq.~\eqref{Hcstar}.  The green dotted zone allows an FF state solution to exist, while it has higher free energy than the Bloch SC state. A commensurate-incommensurate (CI) transition occurs at the phase boundary between the Bloch SC state and the FF state. Two red stars mark Bloch SC states (A) and (B) at $|g|/|\alpha|=1.0$ and A: $(H/H_{\m{c}2\perp}^\m{ML})(a/\xi)=1.8$ and B: $(H/H_{\m{c}2\perp}^\m{ML})(a/\xi)=4.0$, respectively. }\label{phasediag}
\end{figure}

\emph{Phase diagram.}
As discussed above, for a given set of independent parameters $(g/|\alpha|,H)$, Eqs.~\eqref{discreteGL} can be solved numerically and the corresponding free-energy density $\fed_\m{Bloch}$ can be computed subsequently. By comparing  $\fed_\m{Bloch}$ with the free-energy density of the decoupled SC state and the FF state, say, $\fed_\m{D}$ and $\fed_\m{FF}$, we are able to obtain a phase diagram consisting of a decoupled SC state, FF state, and Bloch SC state, as shown in Fig.~\ref{phasediag}.
Notice that if $\{n\beta,a_{1\nu},a_{2\nu}\}$ is a solution to Eqs.~\eqref{discreteGL} for a given pair of $(g/|\alpha|,H)$, then $\{n\beta,e^{i\pi\nu}a_{1\nu},e^{i\pi\nu}a_{2\nu}\}$ will be a physically equivalent solution for $(-g/|\alpha|,H)$, so that the phase diagram can be parametrized by $(|g|/|\alpha|,H)$.
Furthermore, $H$ can be replaced by a dimensionless ratio, $(H/H_{\m{c}2\perp}^\m{ML})\cdot(a/\xi)$.
Here, $H_{\m{c}2\perp}^\m{ML}=\Phi_0(2\pi\xi^2)^{-1}$ is the perpendicular upper critical field of a monolayer SC thin film.

As seen in Fig.~\ref{phasediag}, a larger Josephson coupling energy $|g|$ and smaller applied parallel field $H$ favor the FF state, while Bloch SC states will gain more free energy at smaller $|g|$ and larger $H$.
As $|g|$ decreases and/or $H$ increases, the energy cost in the kinetic term in Eq.~\eqref{freeEne} will be canceled by the phase factor $\e^{i\eta_{l}k_0{}x/2}$ in Eq.~\eqref{ansatz}, resulting in $(\pt_x|\tpsi_l|)^2\to0$ and a more and more spatially uniform distribution of the superfluid density $|\psi_{l}(x)|^2$ [see Figs.~\ref{density18} and \ref{density40}]. 
Along the line of $g=0$, the Bloch SC state restores to the decoupled SC state given in Eq.~\eqref{DCstate} by taking a constant $\tpsi_l=\sqrt{\rho_{\m{s}0}}$ in Eq.~\eqref{ansatz}. 
The green dashed line in Fig.~\ref{phasediag} indicates the $H^*_\m{c}$ given by Eq.~\eqref{Hcstar}, below which an FF state solution exists but has a higher free energy than the Bloch SC state in the green dotted zone, until $H$ decreases further and enters the FF phase marked by the green zone.

It is noted that although the Bloch solution exists mathematically throughout the whole phase diagram, an actual parallel critical field $H_{\m{c}\parallel}$ will be determined physically by the parallel critical field for a monolayer thin film $H_{\m{c}\parallel}^\m{ML}=2\sqrt{3}H_{\m{c}2\perp}^\m{ML}(\xi/d)$~\cite{Tinkham} and/or the Pauli paramagnetic limit $H_{P}$~\cite{Chandrasekhar,Clogston}, where $d$ is the thickness of the thin film. The Bloch SC state will vanish when $H>H_{\m{c}\parallel}$. 

\emph{Phase transition.---} The phase transition between the Bloch SC phase and the FF phase, which can be viewed as the melting of the superfluid density wave, turns out to be of the universality class of the commensurate-incommensurate (CI) transition, and is a second-order transition~\cite{Bak82}~\footnote{In the mean-field level, the fact that the amplitude oscillations of the Bloch state increase as approaching the phase boundary indicates a first-order phase transition. However, this phase transition will become continuous when fluctuations are considered. Physically, around the critical point, the fluctuations are dominated by domain wall fluctuations, and the universality class has been determined and is known as the PT universality class (see Ref.~\cite{Bak82} for more details)}. The effective model for such a transition can be written in terms of the relative phase $\phi(\rm{r})$ between the two layers~\cite{appendix},
\begin{equation}\label{eq:PT}
	H_\m{eff} = \rho_{\m{s}} \int d^2\br\,\bigg\{\frac{\hbar^2}{2m^*}|\bm{\nabla}\phi(\br)|^2 + 2g\cos\left[\phi(\br)-k_0x\right]\bigg\}.
\end{equation}
This is exactly the Pokrovsky-Talapov (PT) model in the context of incommensurate crystals~\cite{PT79}, which was adopted to study double-layer quantum Hall systems~\cite{KYang94,KYang96}. In our case, the FF state corresponds to the commensurate state in the PT model, while the Bloch SC state corresponds to the incommensurate state that breaks the translational symmetry.

\emph{Spatial modulation of superfluid density.}
As mentioned, the $\cal{T}\cal{P}_{z}$ symmetry, $\psi_1(x)=\psi_2(x)^*$, is respected by all the found solutions.  Consequently, the two layers have the same local superfluid density $|\psi_{l}(x)|^2$. For a Bloch SC state, as illustrated in Figs.~\ref{density18} and \ref{density40}, the superfluid density is modulated spatially by the applied magnetic field and manifests a PDW in each layer. Since the local superfluid density is associated with a local superconducting gap, this type of PDW can be probed by a scanning tunneling microscope (STM) [see Fig.~\ref{model}]. It is found that the spatial modulation will be enhanced when a Bloch SC state approaches the phase boundary (see the two states A and B in Fig.~\ref{phasediag}).

\begin{figure}[tb]
	\centering
	\subfigure{
		\label{current18}
		\begin{overpic}[width=\linewidth]{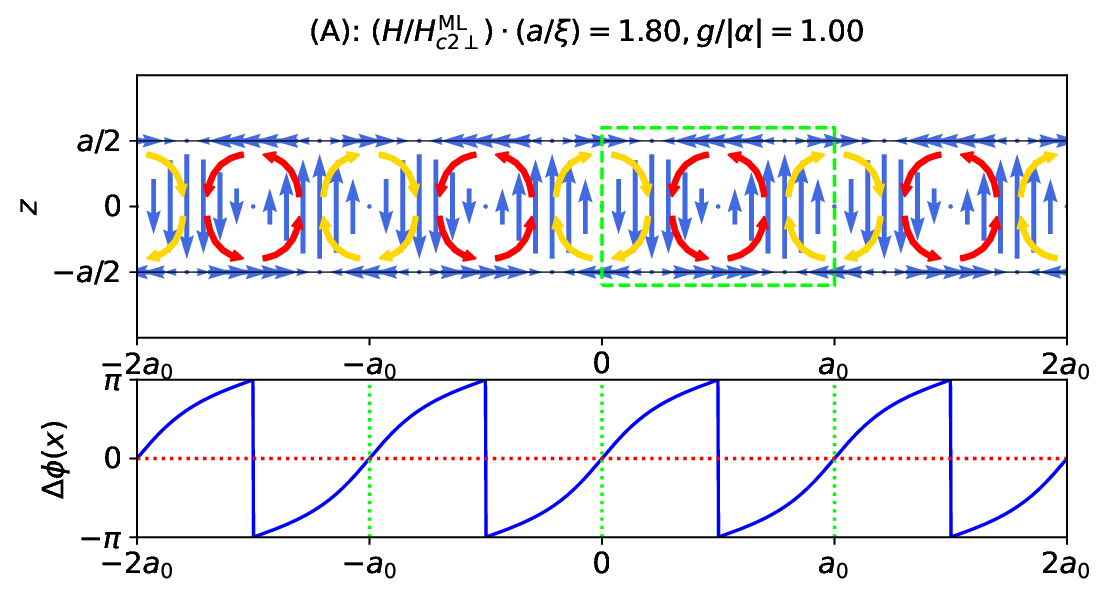}
			\put(0,49){(a)}
		\end{overpic}
	}
	\vfill
	\subfigure{
		\label{current40}
		\begin{overpic}[width=\linewidth]{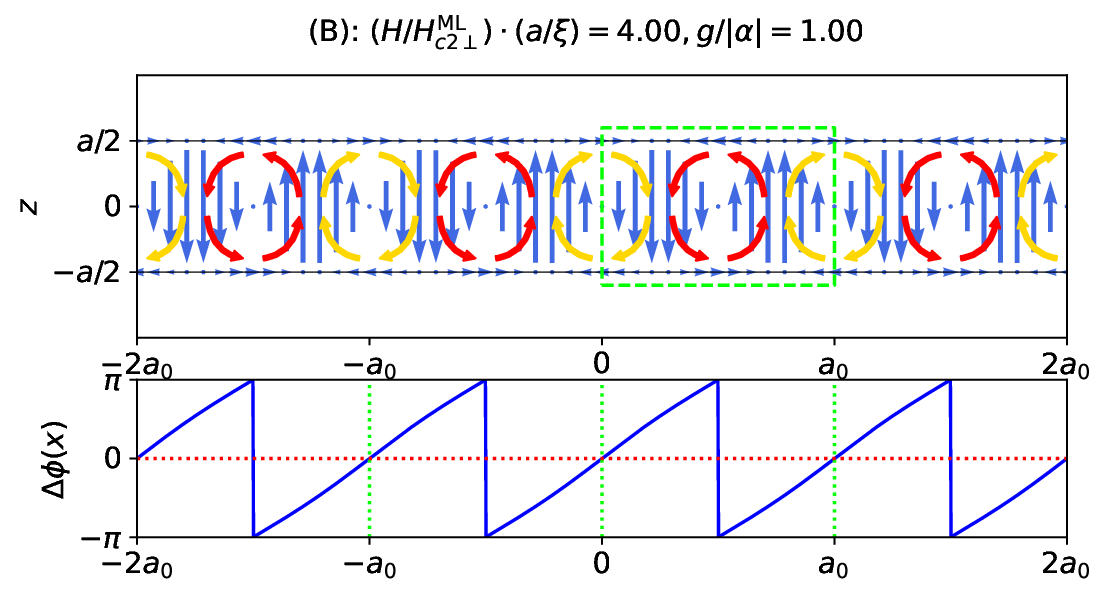}
			\put(0,49){(b)}
		\end{overpic}
	}
	\caption{ The pattern of supercurrent and tunneling current (blue arrows), and the phase difference between the two layers. (a) Bloch SC state A in Fig.~\ref{phasediag}. (b) Bloch SC state B in  Fig.~\ref{phasediag}. Red and yellow arrows indicate Josephson vortices and antivortices. Along each green dashed loop, a London's fluxoid can be defined, which is quantized as $\Phi^{\prime}=\Phi_0$.}
	\label{current}
\end{figure}

\emph{Supercurrent and Josephson tunneling effect.}
Now we proceed to study the intralayer supercurrent and interlayer Josephson tunneling current.
The supercurrent density on each layer reads
$\bf{J}_{\m{s}l} = \frac{e^*\hbar}{2m^*i}\left(\psi_l^*\bm{\nabla}\psi_l-\psi_l\bm{\nabla}\psi_l^*\right) - \frac{e^{*2}}{m^*c}\psi_l^*\psi_l\bf{A}_l$ \cite{Tinkham}, which is in the direction perpendicular to the applied magnetic field $H$ and can be written as $\bf{J}_{\m{s}l} =J_{\m{s}l}(x)\hat{x}$.
On the other hand, the Josephson tunneling effect is characterized by the tunneling current density $J_{\m{T}}(x)=-\frac{e^*a}{\hbar}g\Im\,\Big(\psi_1\psi_2^*\e^{i\frac{2\pi}{\Phi_0}\int_{1}^{2}\bf{A}\cdot d\bf{s}}\Big)$.

For the Bloch SC state given by Eqs.~\eqref{ansatz} and \eqref{expand}, straightforward algebra leads to
\begin{subequations}
	\begin{align}\label{supercurrent}
		J_{\m{s}l} &= n\frac{e^{*2}Ha}{2m^*c}\sum_{\nu\nu'}2\nu a_{l\nu}a_{l\nu'} \cos[(\nu-\nu')k_0x],
	\end{align}
	and
	\begin{equation}\label{tunneling_current}
		J_{\m{T}} = -n\frac{e^*a}{\hbar}g\sum_{\nu\nu'}a_{1\nu}a_{2\nu'} \sin[(\nu-\nu'+1)k_0x].
	\end{equation}
\end{subequations}
Here, the fact that $a_{l\nu}$ are real numbers has been used. Note that both $J_{\m{s}l}$ and $J_{\m{T}}$ are periodic, namely, $J_{\m{s}l}(x+a_{0})=J_{\m{s}l}(x)$ and $J_{\m{T}}(x+a_0)=J_{\m{T}}(x)$. Moreover, owing to the $\cal{T}\cal{P}_{z}$ symmetry, $\psi_{\bar{l}}(x)^{*}=\psi_{l}(x)$, we have $J_{\m{s}\bar{l}}(x)=-J_{\m{s}l}(x)$, i.e., the local supercurrent flows in opposite directions on the two layers.

As examples, two typical Bloch solutions have been found numerically at $|g|/|\alpha|=1.0$ and $(H/H_{\m{c}2\perp}^\m{ML})(a/\xi)=1.8\text{ and } 4.0$ respectively, which correspond to two points marked by the red stars (A and B) in the phase diagram in Fig.~\ref{phasediag}. The supercurrent density and the tunneling current density are plotted in Fig.~\ref{current}. These staggered currents form Josephson vortices and antivortices, as indicated by red and yellow arrows in Fig.~\ref{current}.
It is remarkable that this array of Josephson vortex-antivortex pairs is different from the Josephson vortex lattice in the literature~\cite{Koshelev,Berdiyorov, Curran}, although both are induced by a parallel magnetic field.
The orbital magnetization generated by these currents is staggered, and the net orbital magnetization will vanish. This can be revealed by the London's fluxoid. 

\emph{Fluxoid quantization.} Considering a loop enclosing a period of $a_0$ (see the green dashed loops in Fig.~\ref{current}), the London's fluxoid is defined as $\Phi^{\prime}=\Phi+\frac{m^{*}c}{e^{*2}}\oint\frac{\bf{J}_\m{s}}{\rho_\m{s}}\cdot d\bf{s}$, where  $\Phi=\oint\bf{A}\cdot d\bf{s}=Haa_0=\Phi_0$ is the the ordinary flux. For a Bloch SC state, we have $J_{\m{T}}(x+a_0)=J_{\m{T}}(x)$, and the second part in the fluxoid reads $\frac{m^{*}c}{e^{*2}}\oint\frac{\bf{J}_\m{s}}{\rho_\m{s}}\cdot d\bf{s}=\frac{\Phi_0}{2\pi}\oint\bm{\nabla}\varphi\cdot d\bf{s}-\oint\bf{A}\cdot d\bf{s}$. Therefore, $\Phi^{\prime}=\Phi_0 $ is exact the flux quantum, and the net orbital magnetization is counted by the flux $\frac{m^{*}c}{e^{*2}}\oint\frac{\bf{J}_\m{s}}{\rho_\m{s}}\cdot d\bf{s}=0$ and vanishes.

In contrast, an FF state has uniform supercurrents $J_{\m{s}l}^\m{FF} = -\eta_{l}(e^{*2}Ha/2m^{*}c)\rho_\m{s}^\m{FF}$, which flow in opposite directions on the two layers and are perpendicular to the applied magnetic field. Meanwhile, the Josephson tunneling current vanishes, i.e., $J_{\m{T}}^\m{FF} \propto -g\sin(\De\vp)=0$, because the phase difference between the two layers is $\De\vp=0$ or $\pi$. Thus, for an FF state, $\frac{m^{*}c}{e^{*2}}\oint\frac{\bf{J}_\m{s}}{\rho_\m{s}}\cdot d\bf{s}=-\Phi_0$ and $\Phi^{\prime}=0$, suggesting perfect diamagnetism.

\emph{Approximate analytical solutions.} It is remarkable that the Bloch SC state can be well approximated by the Jacobian elliptic functions $\cn$ and $\sn$. When $g=0$, Eqs.~\eqref{GLeq} reduce to two decoupled nonlinear Schr\"{o}dinger equations, and each of them possesses exact periodic instanton solutions in the form of Jacobian elliptic functions~\cite{periodic_instantons,Funakubo}. While exact solutions are not available at a finite Josephson coupling $g$, we would like to propose approximate solutions as follows,
\begin{equation}\label{trial}
	\psi_l(x) = \left[\mu\cn(ux;r)+i\eta_l\nu\sn(ux;r)\right]\e^{i\vp_l},
\end{equation}
where $\vp_1=\vp_2=0$ for $g>0$ and $\vp_1=-\vp_2=-\pi/2$ for $g<0$. Here, $r\in(0,1)$ is the modulus of the elliptic integral $K(r)$~\footnote{The modulus $r$ characterizes the shape of Jacobian elliptic functions, e.g., $\cn(x;0)=\cos(x),\sn(x;0)=\sin(x)$ and $\cn(x;1)=\m{sech}(x),\sn(x;1)=\tanh(x)$.}, $u$ is determined by the period through $4K(r)=2u{}a_0$, and $\mu$ and $\nu$ are two constants and can be treated as variational parameters, which can be obtained by substituting Eq.~\eqref{trial} into Eq.~\eqref{freeEne} and minimizing the free energy with respect to $\mu$, $\nu$, and $r$.
As examples, the optimization for the two states A and B in Fig.~\ref{phasediag}  gives rise to $r=0.6023(90)$ and $r=0.0559(55)$ for A and B, respectively.

By comparing the optimized solution given in Eq.~\eqref{trial} with that solved from Eqs.~\eqref{discreteGL}, we found that these two agree with each other with amazingly high precision. In a wide range of parameters $(g/|\alpha|,H)$, the difference between the corresponding free energies is less than $0.01\%$~\cite{appendix}.

\emph{Summary and discussions.}
In summary, we have found that two weakly linked SC ultrathin films in an applied parallel magnetic field $\Hpara$ can harbor various FFLO or PDW states, including the usual FF state and the proposed Bloch SC state. The latter is indeed a \emph{superfluid density wave} state, and the spatial modulation of superfluid density can be verified by future STM experiments. The Bloch SC state is also characterized by its supercurrent pattern, which forms staggered loops. Thereby an array of Josephson vortex-antivortex pairs comes into being, instead of the usual Josephson vortex array. The phase transition between the FF state and the Bloch SC state is of second order, and can be described by the PT model~\cite{PT79}.

Finally, the Bloch SC state is robust against an extra perpendicular magnetic field $\Hperp$. The problem of finding out the upper critical magnetic field $H^\m{BL}_{\m{c}2\perp}$ in such a bilayer SC system can be mapped to the Rabi model in quantum optics~\cite{Rabi}, which can be solved numerically~\cite{RabiSol,JCsol}. It turns out that there exists a finite $H^\m{BL}_{\m{c}2\perp}$, which is larger than the monolayer value $H^\m{ML}_{\m{c}2\perp}$ as long as the Josephson coupling $g$ is nonzero~\cite{appendix}.

\emph{Acknowledgments.}
We thank Dong-Hui Xu, Hong Yao and Noah F. Q. Yuan for helpful discussions. In particular, we would like to thank Kun Yang for correcting the label of the FF state and bringing Refs.~\cite{PT79,KYang94,KYang96} to our attention. This work is partially supported by National Natural Science Foundation of China (No. 12034004 and No. 11774306) and the Strategic Priority Research Program of Chinese Academy of Sciences (No. XDB28000000).

\bibliography{reference} 

\begin{thebibliography}{46}%
\makeatletter
\providecommand \@ifxundefined [1]{%
 \@ifx{#1\undefined}
}%
\providecommand \@ifnum [1]{%
 \ifnum #1\expandafter \@firstoftwo
 \else \expandafter \@secondoftwo
 \fi
}%
\providecommand \@ifx [1]{%
 \ifx #1\expandafter \@firstoftwo
 \else \expandafter \@secondoftwo
 \fi
}%
\providecommand \natexlab [1]{#1}%
\providecommand \enquote  [1]{``#1''}%
\providecommand \bibnamefont  [1]{#1}%
\providecommand \bibfnamefont [1]{#1}%
\providecommand \citenamefont [1]{#1}%
\providecommand \href@noop [0]{\@secondoftwo}%
\providecommand \href [0]{\begingroup \@sanitize@url \@href}%
\providecommand \@href[1]{\@@startlink{#1}\@@href}%
\providecommand \@@href[1]{\endgroup#1\@@endlink}%
\providecommand \@sanitize@url [0]{\catcode `\\12\catcode `\$12\catcode
  `\&12\catcode `\#12\catcode `\^12\catcode `\_12\catcode `\%12\relax}%
\providecommand \@@startlink[1]{}%
\providecommand \@@endlink[0]{}%
\providecommand \url  [0]{\begingroup\@sanitize@url \@url }%
\providecommand \@url [1]{\endgroup\@href {#1}{\urlprefix }}%
\providecommand \urlprefix  [0]{URL }%
\providecommand \Eprint [0]{\href }%
\providecommand \doibase [0]{http://dx.doi.org/}%
\providecommand \selectlanguage [0]{\@gobble}%
\providecommand \bibinfo  [0]{\@secondoftwo}%
\providecommand \bibfield  [0]{\@secondoftwo}%
\providecommand \translation [1]{[#1]}%
\providecommand \BibitemOpen [0]{}%
\providecommand \bibitemStop [0]{}%
\providecommand \bibitemNoStop [0]{.\EOS\space}%
\providecommand \EOS [0]{\spacefactor3000\relax}%
\providecommand \BibitemShut  [1]{\csname bibitem#1\endcsname}%
\let\auto@bib@innerbib\@empty
\bibitem [{\citenamefont {Casalbuoni}\ and\ \citenamefont
  {Nardulli}(2004)}]{RMP04}%
  \BibitemOpen
  \bibfield  {author} {\bibinfo {author} {\bibfnamefont {R.}~\bibnamefont
  {Casalbuoni}}\ and\ \bibinfo {author} {\bibfnamefont {G.}~\bibnamefont
  {Nardulli}},\ }\href {\doibase 10.1103/RevModPhys.76.263} {\bibfield
  {journal} {\bibinfo  {journal} {Rev. Mod. Phys.}\ }\textbf {\bibinfo {volume}
  {76}},\ \bibinfo {pages} {263} (\bibinfo {year} {2004})}\BibitemShut
  {NoStop}%
\bibitem [{\citenamefont {Fulde}\ and\ \citenamefont
  {Ferrell}(1964)}]{FFstate}%
  \BibitemOpen
  \bibfield  {author} {\bibinfo {author} {\bibfnamefont {P.}~\bibnamefont
  {Fulde}}\ and\ \bibinfo {author} {\bibfnamefont {R.~A.}\ \bibnamefont
  {Ferrell}},\ }\href {\doibase 10.1103/PhysRev.135.A550} {\bibfield  {journal}
  {\bibinfo  {journal} {Phys. Rev.}\ }\textbf {\bibinfo {volume} {135}},\
  \bibinfo {pages} {A550} (\bibinfo {year} {1964})}\BibitemShut {NoStop}%
\bibitem [{\citenamefont {Larkin}\ and\ \citenamefont
  {Ovchinnikov}(1964)}]{LOstate}%
  \BibitemOpen
  \bibfield  {author} {\bibinfo {author} {\bibfnamefont {A.~I.}\ \bibnamefont
  {Larkin}}\ and\ \bibinfo {author} {\bibfnamefont {Y.}~\bibnamefont
  {Ovchinnikov}},\ }\href@noop {} {\bibfield  {journal} {\bibinfo  {journal}
  {Zh. Eksp. Teor. Fiz.}\ }\textbf {\bibinfo {volume} {47}},\ \bibinfo {pages}
  {1136} (\bibinfo {year} {1964})}\BibitemShut {NoStop}%
\bibitem [{\citenamefont {Agterberg}\ \emph {et~al.}(2020)\citenamefont
  {Agterberg}, \citenamefont {Davis}, \citenamefont {Edkins}, \citenamefont
  {Fradkin}, \citenamefont {Van~Harlingen}, \citenamefont {Kivelson},
  \citenamefont {Lee}, \citenamefont {Radzihovsky}, \citenamefont {Tranquada},\
  and\ \citenamefont {Wang}}]{ARCMP20}%
  \BibitemOpen
  \bibfield  {author} {\bibinfo {author} {\bibfnamefont {D.~F.}\ \bibnamefont
  {Agterberg}}, \bibinfo {author} {\bibfnamefont {J.~S.}\ \bibnamefont
  {Davis}}, \bibinfo {author} {\bibfnamefont {S.~D.}\ \bibnamefont {Edkins}},
  \bibinfo {author} {\bibfnamefont {E.}~\bibnamefont {Fradkin}}, \bibinfo
  {author} {\bibfnamefont {D.~J.}\ \bibnamefont {Van~Harlingen}}, \bibinfo
  {author} {\bibfnamefont {S.~A.}\ \bibnamefont {Kivelson}}, \bibinfo {author}
  {\bibfnamefont {P.~A.}\ \bibnamefont {Lee}}, \bibinfo {author} {\bibfnamefont
  {L.}~\bibnamefont {Radzihovsky}}, \bibinfo {author} {\bibfnamefont {J.~M.}\
  \bibnamefont {Tranquada}}, \ and\ \bibinfo {author} {\bibfnamefont
  {Y.}~\bibnamefont {Wang}},\ }\href {\doibase
  10.1146/annurev-conmatphys-031119-050711} {\bibfield  {journal} {\bibinfo
  {journal} {Annu. Rev. Condens. Matter Phys.}\ }\textbf {\bibinfo {volume}
  {11}},\ \bibinfo {pages} {231} (\bibinfo {year} {2020})}\BibitemShut
  {NoStop}%
\bibitem [{\citenamefont {Eom}\ \emph {et~al.}(2006)\citenamefont {Eom},
  \citenamefont {Qin}, \citenamefont {Chou},\ and\ \citenamefont {Shih}}]{Eom}%
  \BibitemOpen
  \bibfield  {author} {\bibinfo {author} {\bibfnamefont {D.}~\bibnamefont
  {Eom}}, \bibinfo {author} {\bibfnamefont {S.}~\bibnamefont {Qin}}, \bibinfo
  {author} {\bibfnamefont {M.-Y.}\ \bibnamefont {Chou}}, \ and\ \bibinfo
  {author} {\bibfnamefont {C.~K.}\ \bibnamefont {Shih}},\ }\href {\doibase
  10.1103/PhysRevLett.96.027005} {\bibfield  {journal} {\bibinfo  {journal}
  {Phys. Rev. Lett.}\ }\textbf {\bibinfo {volume} {96}},\ \bibinfo {pages}
  {027005} (\bibinfo {year} {2006})}\BibitemShut {NoStop}%
\bibitem [{\citenamefont {Morshedloo}\ \emph {et~al.}(2015)\citenamefont
  {Morshedloo}, \citenamefont {Roknabadi},\ and\ \citenamefont
  {Behdani}}]{Morshedloo}%
  \BibitemOpen
  \bibfield  {author} {\bibinfo {author} {\bibfnamefont {T.}~\bibnamefont
  {Morshedloo}}, \bibinfo {author} {\bibfnamefont {M.}~\bibnamefont
  {Roknabadi}}, \ and\ \bibinfo {author} {\bibfnamefont {M.}~\bibnamefont
  {Behdani}},\ }\href {\doibase https://doi.org/10.1016/j.physc.2014.11.006}
  {\bibfield  {journal} {\bibinfo  {journal} {Phys. C: Supercond. Appl.}\
  }\textbf {\bibinfo {volume} {509}},\ \bibinfo {pages} {1} (\bibinfo {year}
  {2015})}\BibitemShut {NoStop}%
\bibitem [{\citenamefont {Bulaevski{\u{\i}}}(1975)}]{Bulaevski}%
  \BibitemOpen
  \bibfield  {author} {\bibinfo {author} {\bibfnamefont {L.~N.}\ \bibnamefont
  {Bulaevski{\u{\i}}}},\ }\href {\doibase 10.1070/pu1975v018n07abeh004892}
  {\bibfield  {journal} {\bibinfo  {journal} {Sov. Phys. Usp.}\ }\textbf
  {\bibinfo {volume} {18}},\ \bibinfo {pages} {514} (\bibinfo {year}
  {1975})}\BibitemShut {NoStop}%
\bibitem [{\citenamefont {Josephson}(1962)}]{Josephson1962}%
  \BibitemOpen
  \bibfield  {author} {\bibinfo {author} {\bibfnamefont {B.~D.}\ \bibnamefont
  {Josephson}},\ }\href {\doibase 10.1016/0031-9163(62)91369-0} {\bibfield
  {journal} {\bibinfo  {journal} {Phys. Lett.}\ }\textbf {\bibinfo {volume}
  {1}},\ \bibinfo {pages} {251} (\bibinfo {year} {1962})}\BibitemShut {NoStop}%
\bibitem [{\citenamefont {Josephson}(1965)}]{Josephson1965}%
  \BibitemOpen
  \bibfield  {author} {\bibinfo {author} {\bibfnamefont {B.~D.}\ \bibnamefont
  {Josephson}},\ }\href {\doibase 10.1080/00018736500101091} {\bibfield
  {journal} {\bibinfo  {journal} {Adv. Phys.}\ }\textbf {\bibinfo {volume}
  {14}},\ \bibinfo {pages} {419} (\bibinfo {year} {1965})}\BibitemShut
  {NoStop}%
\bibitem [{\citenamefont {Likharev}(1979)}]{RMP79}%
  \BibitemOpen
  \bibfield  {author} {\bibinfo {author} {\bibfnamefont {K.~K.}\ \bibnamefont
  {Likharev}},\ }\href {\doibase 10.1103/RevModPhys.51.101} {\bibfield
  {journal} {\bibinfo  {journal} {Rev. Mod. Phys.}\ }\textbf {\bibinfo {volume}
  {51}},\ \bibinfo {pages} {101} (\bibinfo {year} {1979})}\BibitemShut
  {NoStop}%
\bibitem [{\citenamefont {Kleiner}\ \emph {et~al.}(1992)\citenamefont
  {Kleiner}, \citenamefont {Steinmeyer}, \citenamefont {Kunkel},\ and\
  \citenamefont {M\"uller}}]{Kleiner1992}%
  \BibitemOpen
  \bibfield  {author} {\bibinfo {author} {\bibfnamefont {R.}~\bibnamefont
  {Kleiner}}, \bibinfo {author} {\bibfnamefont {F.}~\bibnamefont {Steinmeyer}},
  \bibinfo {author} {\bibfnamefont {G.}~\bibnamefont {Kunkel}}, \ and\ \bibinfo
  {author} {\bibfnamefont {P.}~\bibnamefont {M\"uller}},\ }\href {\doibase
  10.1103/PhysRevLett.68.2394} {\bibfield  {journal} {\bibinfo  {journal}
  {Phys. Rev. Lett.}\ }\textbf {\bibinfo {volume} {68}},\ \bibinfo {pages}
  {2394} (\bibinfo {year} {1992})}\BibitemShut {NoStop}%
\bibitem [{\citenamefont {Kleiner}\ and\ \citenamefont
  {M\"uller}(1994)}]{Kleiner1994}%
  \BibitemOpen
  \bibfield  {author} {\bibinfo {author} {\bibfnamefont {R.}~\bibnamefont
  {Kleiner}}\ and\ \bibinfo {author} {\bibfnamefont {P.}~\bibnamefont
  {M\"uller}},\ }\href {\doibase 10.1103/PhysRevB.49.1327} {\bibfield
  {journal} {\bibinfo  {journal} {Phys. Rev. B}\ }\textbf {\bibinfo {volume}
  {49}},\ \bibinfo {pages} {1327} (\bibinfo {year} {1994})}\BibitemShut
  {NoStop}%
\bibitem [{\citenamefont {Agterberg}(2003)}]{Agterberg}%
  \BibitemOpen
  \bibfield  {author} {\bibinfo {author} {\bibfnamefont {D.}~\bibnamefont
  {Agterberg}},\ }\href {\doibase
  https://doi.org/10.1016/S0921-4534(03)00634-8} {\bibfield  {journal}
  {\bibinfo  {journal} {Phys. C: Supercond. Appl.}\ }\textbf {\bibinfo {volume}
  {387}},\ \bibinfo {pages} {13} (\bibinfo {year} {2003})}\BibitemShut
  {NoStop}%
\bibitem [{\citenamefont {Barzykin}\ and\ \citenamefont
  {Gor'kov}(2002)}]{Barzykin}%
  \BibitemOpen
  \bibfield  {author} {\bibinfo {author} {\bibfnamefont {V.}~\bibnamefont
  {Barzykin}}\ and\ \bibinfo {author} {\bibfnamefont {L.~P.}\ \bibnamefont
  {Gor'kov}},\ }\href {\doibase 10.1103/PhysRevLett.89.227002} {\bibfield
  {journal} {\bibinfo  {journal} {Phys. Rev. Lett.}\ }\textbf {\bibinfo
  {volume} {89}},\ \bibinfo {pages} {227002} (\bibinfo {year}
  {2002})}\BibitemShut {NoStop}%
\bibitem [{\citenamefont {Aoyama}\ and\ \citenamefont
  {Sigrist}(2012)}]{Aoyama}%
  \BibitemOpen
  \bibfield  {author} {\bibinfo {author} {\bibfnamefont {K.}~\bibnamefont
  {Aoyama}}\ and\ \bibinfo {author} {\bibfnamefont {M.}~\bibnamefont
  {Sigrist}},\ }\href {\doibase 10.1103/PhysRevLett.109.237007} {\bibfield
  {journal} {\bibinfo  {journal} {Phys. Rev. Lett.}\ }\textbf {\bibinfo
  {volume} {109}},\ \bibinfo {pages} {237007} (\bibinfo {year}
  {2012})}\BibitemShut {NoStop}%
\bibitem [{\citenamefont {Dimitrova}\ and\ \citenamefont
  {Feigel'man}(2003)}]{Dimitrova}%
  \BibitemOpen
  \bibfield  {author} {\bibinfo {author} {\bibfnamefont {O.~V.}\ \bibnamefont
  {Dimitrova}}\ and\ \bibinfo {author} {\bibfnamefont {M.~V.}\ \bibnamefont
  {Feigel'man}},\ }\href {\doibase 10.1134/1.1644308} {\bibfield  {journal}
  {\bibinfo  {journal} {J. Exp. Theor. Phys. Lett.}\ }\textbf {\bibinfo
  {volume} {78}},\ \bibinfo {pages} {637} (\bibinfo {year} {2003})}\BibitemShut
  {NoStop}%
\bibitem [{\citenamefont {Liu}(2017)}]{bilayerTMD}%
  \BibitemOpen
  \bibfield  {author} {\bibinfo {author} {\bibfnamefont {C.-X.}\ \bibnamefont
  {Liu}},\ }\href {\doibase 10.1103/PhysRevLett.118.087001} {\bibfield
  {journal} {\bibinfo  {journal} {Phys. Rev. Lett.}\ }\textbf {\bibinfo
  {volume} {118}},\ \bibinfo {pages} {087001} (\bibinfo {year}
  {2017})}\BibitemShut {NoStop}%
\bibitem [{\citenamefont {Lu}\ \emph {et~al.}(2015)\citenamefont {Lu},
  \citenamefont {Zheliuk}, \citenamefont {Leermakers}, \citenamefont {Yuan},
  \citenamefont {Zeitler}, \citenamefont {Law},\ and\ \citenamefont {Ye}}]{Lu}%
  \BibitemOpen
  \bibfield  {author} {\bibinfo {author} {\bibfnamefont {J.~M.}\ \bibnamefont
  {Lu}}, \bibinfo {author} {\bibfnamefont {O.}~\bibnamefont {Zheliuk}},
  \bibinfo {author} {\bibfnamefont {I.}~\bibnamefont {Leermakers}}, \bibinfo
  {author} {\bibfnamefont {N.~F.~Q.}\ \bibnamefont {Yuan}}, \bibinfo {author}
  {\bibfnamefont {U.}~\bibnamefont {Zeitler}}, \bibinfo {author} {\bibfnamefont
  {K.~T.}\ \bibnamefont {Law}}, \ and\ \bibinfo {author} {\bibfnamefont
  {J.~T.}\ \bibnamefont {Ye}},\ }\href {\doibase 10.1126/science.aab2277}
  {\bibfield  {journal} {\bibinfo  {journal} {Science}\ }\textbf {\bibinfo
  {volume} {350}},\ \bibinfo {pages} {1353} (\bibinfo {year}
  {2015})}\BibitemShut {NoStop}%
\bibitem [{\citenamefont {Saito}\ \emph {et~al.}(2016)\citenamefont {Saito},
  \citenamefont {Nakamura}, \citenamefont {Bahramy}, \citenamefont {Kohama},
  \citenamefont {Ye}, \citenamefont {Kasahara}, \citenamefont {Nakagawa},
  \citenamefont {Onga}, \citenamefont {Tokunaga}, \citenamefont {Nojima},
  \citenamefont {Yanase},\ and\ \citenamefont {Iwasa}}]{Yu}%
  \BibitemOpen
  \bibfield  {author} {\bibinfo {author} {\bibfnamefont {Y.}~\bibnamefont
  {Saito}}, \bibinfo {author} {\bibfnamefont {Y.}~\bibnamefont {Nakamura}},
  \bibinfo {author} {\bibfnamefont {M.~S.}\ \bibnamefont {Bahramy}}, \bibinfo
  {author} {\bibfnamefont {Y.}~\bibnamefont {Kohama}}, \bibinfo {author}
  {\bibfnamefont {J.}~\bibnamefont {Ye}}, \bibinfo {author} {\bibfnamefont
  {Y.}~\bibnamefont {Kasahara}}, \bibinfo {author} {\bibfnamefont
  {Y.}~\bibnamefont {Nakagawa}}, \bibinfo {author} {\bibfnamefont
  {M.}~\bibnamefont {Onga}}, \bibinfo {author} {\bibfnamefont {M.}~\bibnamefont
  {Tokunaga}}, \bibinfo {author} {\bibfnamefont {T.}~\bibnamefont {Nojima}},
  \bibinfo {author} {\bibfnamefont {Y.}~\bibnamefont {Yanase}}, \ and\ \bibinfo
  {author} {\bibfnamefont {Y.}~\bibnamefont {Iwasa}},\ }\href {\doibase
  10.1038/nphys3580} {\bibfield  {journal} {\bibinfo  {journal} {Nat. Phys.}\
  }\textbf {\bibinfo {volume} {12}},\ \bibinfo {pages} {144} (\bibinfo {year}
  {2016})}\BibitemShut {NoStop}%
\bibitem [{\citenamefont {Xi}\ \emph {et~al.}(2016)\citenamefont {Xi},
  \citenamefont {Wang}, \citenamefont {Zhao}, \citenamefont {Park},
  \citenamefont {Law}, \citenamefont {Berger}, \citenamefont {Forró},
  \citenamefont {Shan},\ and\ \citenamefont {Mak}}]{Xi}%
  \BibitemOpen
  \bibfield  {author} {\bibinfo {author} {\bibfnamefont {X.}~\bibnamefont
  {Xi}}, \bibinfo {author} {\bibfnamefont {Z.}~\bibnamefont {Wang}}, \bibinfo
  {author} {\bibfnamefont {W.}~\bibnamefont {Zhao}}, \bibinfo {author}
  {\bibfnamefont {J.-H.}\ \bibnamefont {Park}}, \bibinfo {author}
  {\bibfnamefont {K.~T.}\ \bibnamefont {Law}}, \bibinfo {author} {\bibfnamefont
  {H.}~\bibnamefont {Berger}}, \bibinfo {author} {\bibfnamefont
  {L.}~\bibnamefont {Forró}}, \bibinfo {author} {\bibfnamefont
  {J.}~\bibnamefont {Shan}}, \ and\ \bibinfo {author} {\bibfnamefont {K.~F.}\
  \bibnamefont {Mak}},\ }\href {\doibase 10.1038/nphys3538} {\bibfield
  {journal} {\bibinfo  {journal} {Nat. Phys.}\ }\textbf {\bibinfo {volume}
  {12}},\ \bibinfo {pages} {139} (\bibinfo {year} {2016})}\BibitemShut
  {NoStop}%
\bibitem [{\citenamefont {Zhu}\ \emph {et~al.}(2021)\citenamefont {Zhu},
  \citenamefont {Papaj}, \citenamefont {Nie}, \citenamefont {Xu}, \citenamefont
  {Gu}, \citenamefont {Yang}, \citenamefont {Guan}, \citenamefont {Wang},
  \citenamefont {Li}, \citenamefont {Liu}, \citenamefont {Luo}, \citenamefont
  {Xu}, \citenamefont {Zheng}, \citenamefont {Fu},\ and\ \citenamefont
  {Jia}}]{Zheng21}%
  \BibitemOpen
  \bibfield  {author} {\bibinfo {author} {\bibfnamefont {Z.}~\bibnamefont
  {Zhu}}, \bibinfo {author} {\bibfnamefont {M.}~\bibnamefont {Papaj}}, \bibinfo
  {author} {\bibfnamefont {X.-A.}\ \bibnamefont {Nie}}, \bibinfo {author}
  {\bibfnamefont {H.-K.}\ \bibnamefont {Xu}}, \bibinfo {author} {\bibfnamefont
  {Y.-S.}\ \bibnamefont {Gu}}, \bibinfo {author} {\bibfnamefont
  {X.}~\bibnamefont {Yang}}, \bibinfo {author} {\bibfnamefont {D.}~\bibnamefont
  {Guan}}, \bibinfo {author} {\bibfnamefont {S.}~\bibnamefont {Wang}}, \bibinfo
  {author} {\bibfnamefont {Y.}~\bibnamefont {Li}}, \bibinfo {author}
  {\bibfnamefont {C.}~\bibnamefont {Liu}}, \bibinfo {author} {\bibfnamefont
  {J.}~\bibnamefont {Luo}}, \bibinfo {author} {\bibfnamefont {Z.-A.}\
  \bibnamefont {Xu}}, \bibinfo {author} {\bibfnamefont {H.}~\bibnamefont
  {Zheng}}, \bibinfo {author} {\bibfnamefont {L.}~\bibnamefont {Fu}}, \ and\
  \bibinfo {author} {\bibfnamefont {J.-F.}\ \bibnamefont {Jia}},\ }\href
  {\doibase 10.1126/science.abf1077} {\bibfield  {journal} {\bibinfo  {journal}
  {Science}\ }\textbf {\bibinfo {volume} {374}},\ \bibinfo {pages} {1381}
  (\bibinfo {year} {2021})}\BibitemShut {NoStop}%
\bibitem [{\citenamefont {Lawrence}\ and\ \citenamefont
  {Doniach}(1971)}]{LawrenceDoniach}%
  \BibitemOpen
  \bibfield  {author} {\bibinfo {author} {\bibfnamefont {W.~E.}\ \bibnamefont
  {Lawrence}}\ and\ \bibinfo {author} {\bibfnamefont {S.}~\bibnamefont
  {Doniach}},\ }in\ \href@noop {} {\emph {\bibinfo {booktitle} {Proceedings of
  the 12th International Conference on Low Temperature Physics}}},\ \bibinfo
  {editor} {edited by\ \bibinfo {editor} {\bibfnamefont {E.}~\bibnamefont
  {Kanda}}}\ (\bibinfo  {publisher} {Academic},\ \bibinfo {address} {Kyoto},\
  \bibinfo {year} {1971})\ p.\ \bibinfo {pages} {361}\BibitemShut {NoStop}%
\bibitem [{\citenamefont {Yang}\ and\ \citenamefont {Sondhi}(2000)}]{YangKun}%
  \BibitemOpen
  \bibfield  {author} {\bibinfo {author} {\bibfnamefont {K.}~\bibnamefont
  {Yang}}\ and\ \bibinfo {author} {\bibfnamefont {S.~L.}\ \bibnamefont
  {Sondhi}},\ }\href {\doibase 10.1063/1.373400} {\bibfield  {journal}
  {\bibinfo  {journal} {Journal of Applied Physics}\ }\textbf {\bibinfo
  {volume} {87}},\ \bibinfo {pages} {5549} (\bibinfo {year}
  {2000})}\BibitemShut {NoStop}%
\bibitem [{\citenamefont {Ginzburg}\ and\ \citenamefont
  {Landau}(1950)}]{Ginzburg-Landau}%
  \BibitemOpen
  \bibfield  {author} {\bibinfo {author} {\bibfnamefont {V.~L.}\ \bibnamefont
  {Ginzburg}}\ and\ \bibinfo {author} {\bibfnamefont {L.~D.}\ \bibnamefont
  {Landau}},\ }\href@noop {} {\bibfield  {journal} {\bibinfo  {journal} {Zh.
  Eksp. Teor. Fiz.}\ }\textbf {\bibinfo {volume} {20}},\ \bibinfo {pages}
  {1064} (\bibinfo {year} {1950})}\BibitemShut {NoStop}%
\bibitem [{Note1()}]{Note1}%
  \BibitemOpen
  \bibinfo {note} {At first sight, the SC order parameter $\psi _{l}(x) $ does
  not vary spatially and suggests a uniform state rather than an FF state,
  which is true in the absence of a magnetic field. However, the situation
  changes in the presence of a magnetic field, because the SC order parameter
  does depend on the gauge choice and we should consider the gauge invariant
  physical observable, say, the current, instead. Indeed, this state carries
  finite and opposite supercurrents in both layers, so that it is an FF state
  by definition.}\BibitemShut {Stop}%
\bibitem [{\citenamefont {Machholm}\ \emph {et~al.}(2003)\citenamefont
  {Machholm}, \citenamefont {Pethick},\ and\ \citenamefont {Smith}}]{Pethick}%
  \BibitemOpen
  \bibfield  {author} {\bibinfo {author} {\bibfnamefont {M.}~\bibnamefont
  {Machholm}}, \bibinfo {author} {\bibfnamefont {C.~J.}\ \bibnamefont
  {Pethick}}, \ and\ \bibinfo {author} {\bibfnamefont {H.}~\bibnamefont
  {Smith}},\ }\href {\doibase 10.1103/PhysRevA.67.053613} {\bibfield  {journal}
  {\bibinfo  {journal} {Phys. Rev. A}\ }\textbf {\bibinfo {volume} {67}},\
  \bibinfo {pages} {053613} (\bibinfo {year} {2003})}\BibitemShut {NoStop}%
\bibitem [{app()}]{appendix}%
  \BibitemOpen
  \href@noop {} {\bibinfo  {journal} {See Supplemental Material for more
  details on numerical solution for Bloch SC states, analytical approximation,
  effective model near the commensuration-incommensurate phase transition, and
  the stability of this bilayer SC system against an extra perpendicular
  magnetic field.}\ }\BibitemShut {NoStop}%
\bibitem [{Note2()}]{Note2}%
  \BibitemOpen
\bibfield  {journal} {  }\bibinfo {note} {Because $\phi _{l\nu }$'s occur in a
  free-energy functional in the form of $\protect \qopname \relax o{cos}(\phi
  _{l_1\nu _1}-\phi _{l_2\nu _2})$ and $\protect \qopname \relax o{cos}(\phi
  _{l_1\nu _1}-\phi _{l_2\nu _2}+\phi _{l_3\nu _3}-\phi _{l_4\nu _4})$, and a
  saddle point will be achieved at $\protect \qopname \relax o{cos}(\phi
  _{l_1\nu _1}-\phi _{l_2\nu _2})=\pm {}1$ and $\protect \qopname \relax
  o{cos}(\phi _{l_1\nu _1}-\phi _{l_2\nu _2}+\phi _{l_3\nu _3}-\phi _{l_4\nu
  _4})=\pm {1}$~\cite {Pethick}, so that we can always choose $a_{l\nu }$ as
  real numbers.}\BibitemShut {Stop}%
\bibitem [{\citenamefont {Tinkham}(1996)}]{Tinkham}%
  \BibitemOpen
  \bibfield  {author} {\bibinfo {author} {\bibfnamefont {M.}~\bibnamefont
  {Tinkham}},\ }\href@noop {} {\emph {\bibinfo {title} {Introduction to
  Superconductivity}}},\ \bibinfo {edition} {2nd}\ ed.\ (\bibinfo  {publisher}
  {McGraw-Hill},\ \bibinfo {address} {New York},\ \bibinfo {year}
  {1996})\BibitemShut {NoStop}%
\bibitem [{\citenamefont {Chandrasekhar}(1962)}]{Chandrasekhar}%
  \BibitemOpen
  \bibfield  {author} {\bibinfo {author} {\bibfnamefont {B.~S.}\ \bibnamefont
  {Chandrasekhar}},\ }\href {\doibase 10.1063/1.1777362} {\bibfield  {journal}
  {\bibinfo  {journal} {Appl. Phys. Lett.}\ }\textbf {\bibinfo {volume} {1}},\
  \bibinfo {pages} {7} (\bibinfo {year} {1962})}\BibitemShut {NoStop}%
\bibitem [{\citenamefont {Clogston}(1962)}]{Clogston}%
  \BibitemOpen
  \bibfield  {author} {\bibinfo {author} {\bibfnamefont {A.~M.}\ \bibnamefont
  {Clogston}},\ }\href {\doibase 10.1103/PhysRevLett.9.266} {\bibfield
  {journal} {\bibinfo  {journal} {Phys. Rev. Lett.}\ }\textbf {\bibinfo
  {volume} {9}},\ \bibinfo {pages} {266} (\bibinfo {year} {1962})}\BibitemShut
  {NoStop}%
\bibitem [{\citenamefont {Bak}(1982)}]{Bak82}%
  \BibitemOpen
  \bibfield  {author} {\bibinfo {author} {\bibfnamefont {P.}~\bibnamefont
  {Bak}},\ }\href {\doibase 10.1088/0034-4885/45/6/001} {\bibfield  {journal}
  {\bibinfo  {journal} {Reports on Progress in Physics}\ }\textbf {\bibinfo
  {volume} {45}},\ \bibinfo {pages} {587} (\bibinfo {year} {1982})}\BibitemShut
  {NoStop}%
\bibitem [{Note3()}]{Note3}%
  \BibitemOpen
  \bibinfo {note} {In the mean-field level, the fact that the amplitude
  oscillations of the Bloch state increase as approaching the phase boundary
  indicates a first-order phase transition. However, this phase transition will
  become continuous when fluctuations are considered. Physically, around the
  critical point, the fluctuations are dominated by domain wall fluctuations,
  and the universality class has been determined and is known as the PT
  universality class (see Ref.~\cite {Bak82} for more details)}\BibitemShut
  {NoStop}%
\bibitem [{\citenamefont {Pokrovsky}\ and\ \citenamefont
  {Talapov}(1979)}]{PT79}%
  \BibitemOpen
  \bibfield  {author} {\bibinfo {author} {\bibfnamefont {V.~L.}\ \bibnamefont
  {Pokrovsky}}\ and\ \bibinfo {author} {\bibfnamefont {A.~L.}\ \bibnamefont
  {Talapov}},\ }\href {\doibase 10.1103/PhysRevLett.42.65} {\bibfield
  {journal} {\bibinfo  {journal} {Phys. Rev. Lett.}\ }\textbf {\bibinfo
  {volume} {42}},\ \bibinfo {pages} {65} (\bibinfo {year} {1979})}\BibitemShut
  {NoStop}%
\bibitem [{\citenamefont {Yang}\ \emph {et~al.}(1994)\citenamefont {Yang},
  \citenamefont {Moon}, \citenamefont {Zheng}, \citenamefont {MacDonald},
  \citenamefont {Girvin}, \citenamefont {Yoshioka},\ and\ \citenamefont
  {Zhang}}]{KYang94}%
  \BibitemOpen
  \bibfield  {author} {\bibinfo {author} {\bibfnamefont {K.}~\bibnamefont
  {Yang}}, \bibinfo {author} {\bibfnamefont {K.}~\bibnamefont {Moon}}, \bibinfo
  {author} {\bibfnamefont {L.}~\bibnamefont {Zheng}}, \bibinfo {author}
  {\bibfnamefont {A.~H.}\ \bibnamefont {MacDonald}}, \bibinfo {author}
  {\bibfnamefont {S.~M.}\ \bibnamefont {Girvin}}, \bibinfo {author}
  {\bibfnamefont {D.}~\bibnamefont {Yoshioka}}, \ and\ \bibinfo {author}
  {\bibfnamefont {S.-C.}\ \bibnamefont {Zhang}},\ }\href {\doibase
  10.1103/PhysRevLett.72.732} {\bibfield  {journal} {\bibinfo  {journal} {Phys.
  Rev. Lett.}\ }\textbf {\bibinfo {volume} {72}},\ \bibinfo {pages} {732}
  (\bibinfo {year} {1994})}\BibitemShut {NoStop}%
\bibitem [{\citenamefont {Yang}\ \emph {et~al.}(1996)\citenamefont {Yang},
  \citenamefont {Moon}, \citenamefont {Belkhir}, \citenamefont {Mori},
  \citenamefont {Girvin}, \citenamefont {MacDonald}, \citenamefont {Zheng},\
  and\ \citenamefont {Yoshioka}}]{KYang96}%
  \BibitemOpen
  \bibfield  {author} {\bibinfo {author} {\bibfnamefont {K.}~\bibnamefont
  {Yang}}, \bibinfo {author} {\bibfnamefont {K.}~\bibnamefont {Moon}}, \bibinfo
  {author} {\bibfnamefont {L.}~\bibnamefont {Belkhir}}, \bibinfo {author}
  {\bibfnamefont {H.}~\bibnamefont {Mori}}, \bibinfo {author} {\bibfnamefont
  {S.~M.}\ \bibnamefont {Girvin}}, \bibinfo {author} {\bibfnamefont {A.~H.}\
  \bibnamefont {MacDonald}}, \bibinfo {author} {\bibfnamefont {L.}~\bibnamefont
  {Zheng}}, \ and\ \bibinfo {author} {\bibfnamefont {D.}~\bibnamefont
  {Yoshioka}},\ }\href {\doibase 10.1103/PhysRevB.54.11644} {\bibfield
  {journal} {\bibinfo  {journal} {Phys. Rev. B}\ }\textbf {\bibinfo {volume}
  {54}},\ \bibinfo {pages} {11644} (\bibinfo {year} {1996})}\BibitemShut
  {NoStop}%
\bibitem [{\citenamefont {Koshelev}\ and\ \citenamefont
  {Dodgson}(2013)}]{Koshelev}%
  \BibitemOpen
  \bibfield  {author} {\bibinfo {author} {\bibfnamefont {A.~E.}\ \bibnamefont
  {Koshelev}}\ and\ \bibinfo {author} {\bibfnamefont {M.~J.~W.}\ \bibnamefont
  {Dodgson}},\ }\href {https://doi.org/10.1134/s1063776113110125} {\bibfield
  {journal} {\bibinfo  {journal} {J. Exp. Theor. Phys.}\ }\textbf {\bibinfo
  {volume} {117}},\ \bibinfo {pages} {449} (\bibinfo {year}
  {2013})}\BibitemShut {NoStop}%
\bibitem [{\citenamefont {Berdiyorov}\ \emph {et~al.}(2018)\citenamefont
  {Berdiyorov}, \citenamefont {Milo\u{s}evi\'{c}}, \citenamefont {Kusmartsev},
  \citenamefont {Peeters},\ and\ \citenamefont {Savel'ev}}]{Berdiyorov}%
  \BibitemOpen
  \bibfield  {author} {\bibinfo {author} {\bibfnamefont {G.~R.}\ \bibnamefont
  {Berdiyorov}}, \bibinfo {author} {\bibfnamefont {M.~V.}\ \bibnamefont
  {Milo\u{s}evi\'{c}}}, \bibinfo {author} {\bibfnamefont {F.}~\bibnamefont
  {Kusmartsev}}, \bibinfo {author} {\bibfnamefont {F.~M.}\ \bibnamefont
  {Peeters}}, \ and\ \bibinfo {author} {\bibfnamefont {S.}~\bibnamefont
  {Savel'ev}},\ }\href {https://doi.org/10.1038/s41598-018-21015-7} {\bibfield
  {journal} {\bibinfo  {journal} {Sci. Rep.}\ }\textbf {\bibinfo {volume}
  {8}},\ \bibinfo {pages} {2733} (\bibinfo {year} {2018})}\BibitemShut
  {NoStop}%
\bibitem [{\citenamefont {Curran}\ \emph {et~al.}(2018)\citenamefont {Curran},
  \citenamefont {Mohammed}, \citenamefont {Bending}, \citenamefont {Koshelev},
  \citenamefont {Tsuchiya},\ and\ \citenamefont {Tamegai}}]{Curran}%
  \BibitemOpen
  \bibfield  {author} {\bibinfo {author} {\bibfnamefont {P.~J.}\ \bibnamefont
  {Curran}}, \bibinfo {author} {\bibfnamefont {H.~A.}\ \bibnamefont
  {Mohammed}}, \bibinfo {author} {\bibfnamefont {S.~J.}\ \bibnamefont
  {Bending}}, \bibinfo {author} {\bibfnamefont {A.~E.}\ \bibnamefont
  {Koshelev}}, \bibinfo {author} {\bibfnamefont {Y.}~\bibnamefont {Tsuchiya}},
  \ and\ \bibinfo {author} {\bibfnamefont {T.}~\bibnamefont {Tamegai}},\ }\href
  {https://doi.org/10.1038/s41598-018-28681-7} {\bibfield  {journal} {\bibinfo
  {journal} {Sci. Rep.}\ }\textbf {\bibinfo {volume} {8}},\ \bibinfo {pages}
  {10914} (\bibinfo {year} {2018})}\BibitemShut {NoStop}%
\bibitem [{\citenamefont {Liang}\ and\ \citenamefont
  {M\"uller-Kirsten}(1992)}]{periodic_instantons}%
  \BibitemOpen
  \bibfield  {author} {\bibinfo {author} {\bibfnamefont {J.-Q.}\ \bibnamefont
  {Liang}}\ and\ \bibinfo {author} {\bibfnamefont {H.~J.~W.}\ \bibnamefont
  {M\"uller-Kirsten}},\ }\href {\doibase 10.1103/PhysRevD.46.4685} {\bibfield
  {journal} {\bibinfo  {journal} {Phys. Rev. D}\ }\textbf {\bibinfo {volume}
  {46}},\ \bibinfo {pages} {4685} (\bibinfo {year} {1992})}\BibitemShut
  {NoStop}%
\bibitem [{\citenamefont {Funakubo}\ \emph {et~al.}(1992)\citenamefont
  {Funakubo}, \citenamefont {Otsuki}, \citenamefont {Takenaga},\ and\
  \citenamefont {Toyoda}}]{Funakubo}%
  \BibitemOpen
  \bibfield  {author} {\bibinfo {author} {\bibfnamefont {K.}~\bibnamefont
  {Funakubo}}, \bibinfo {author} {\bibfnamefont {S.}~\bibnamefont {Otsuki}},
  \bibinfo {author} {\bibfnamefont {K.}~\bibnamefont {Takenaga}}, \ and\
  \bibinfo {author} {\bibfnamefont {F.}~\bibnamefont {Toyoda}},\ }\href
  {\doibase 10.1143/ptp/87.3.663} {\bibfield  {journal} {\bibinfo  {journal}
  {Prog. Theor. Phys.}\ }\textbf {\bibinfo {volume} {87}},\ \bibinfo {pages}
  {663} (\bibinfo {year} {1992})}\BibitemShut {NoStop}%
\bibitem [{Note4()}]{Note4}%
  \BibitemOpen
  \bibinfo {note} {The modulus $r$ characterizes the shape of Jacobian elliptic
  functions, e.g., $\protect \tmspace +\thinmuskip {.1667em}\protect \mathrm
  {cn}(x;0)=\protect \qopname \relax o{cos}(x),\protect \tmspace +\thinmuskip
  {.1667em}\protect \mathrm {sn}(x;0)=\protect \qopname \relax o{sin}(x)$ and
  $\protect \tmspace +\thinmuskip {.1667em}\protect \mathrm {cn}(x;1)=\protect
  \mathrm {sech}(x),\protect \tmspace +\thinmuskip {.1667em}\protect \mathrm
  {sn}(x;1)=\protect \qopname \relax o{tanh}(x)$.}\BibitemShut {Stop}%
\bibitem [{\citenamefont {Rabi}(1936)}]{Rabi}%
  \BibitemOpen
  \bibfield  {author} {\bibinfo {author} {\bibfnamefont {I.~I.}\ \bibnamefont
  {Rabi}},\ }\href {\doibase 10.1103/PhysRev.49.324} {\bibfield  {journal}
  {\bibinfo  {journal} {Phys. Rev.}\ }\textbf {\bibinfo {volume} {49}},\
  \bibinfo {pages} {324} (\bibinfo {year} {1936})}\BibitemShut {NoStop}%
\bibitem [{\citenamefont {Braak}(2011)}]{RabiSol}%
  \BibitemOpen
  \bibfield  {author} {\bibinfo {author} {\bibfnamefont {D.}~\bibnamefont
  {Braak}},\ }\href {\doibase 10.1103/PhysRevLett.107.100401} {\bibfield
  {journal} {\bibinfo  {journal} {Phys. Rev. Lett.}\ }\textbf {\bibinfo
  {volume} {107}},\ \bibinfo {pages} {100401} (\bibinfo {year}
  {2011})}\BibitemShut {NoStop}%
\bibitem [{\citenamefont {Chen}\ \emph {et~al.}(2011)\citenamefont {Chen},
  \citenamefont {Liu}, \citenamefont {Zhang},\ and\ \citenamefont
  {Wang}}]{JCsol}%
  \BibitemOpen
  \bibfield  {author} {\bibinfo {author} {\bibfnamefont {Q.-H.}\ \bibnamefont
  {Chen}}, \bibinfo {author} {\bibfnamefont {T.}~\bibnamefont {Liu}}, \bibinfo
  {author} {\bibfnamefont {Y.-Y.}\ \bibnamefont {Zhang}}, \ and\ \bibinfo
  {author} {\bibfnamefont {K.-L.}\ \bibnamefont {Wang}},\ }\href {\doibase
  10.1209/0295-5075/96/14003} {\bibfield  {journal} {\bibinfo  {journal}
  {Europhys. Lett.}\ }\textbf {\bibinfo {volume} {96}},\ \bibinfo {pages}
  {14003} (\bibinfo {year} {2011})}\BibitemShut {NoStop}%
\bibitem [{\citenamefont {Jaynes}\ and\ \citenamefont
  {Cummings}(1963)}]{JCmodel}%
  \BibitemOpen
  \bibfield  {author} {\bibinfo {author} {\bibfnamefont {E.}~\bibnamefont
  {Jaynes}}\ and\ \bibinfo {author} {\bibfnamefont {F.}~\bibnamefont
  {Cummings}},\ }\href {\doibase 10.1109/PROC.1963.1664} {\bibfield  {journal}
  {\bibinfo  {journal} {Proc. IEEE}\ }\textbf {\bibinfo {volume} {51}},\
  \bibinfo {pages} {89} (\bibinfo {year} {1963})}\BibitemShut {NoStop}%
\end{thebibliography}%
\clearpage

\renewcommand{\thetable}{S\arabic{table}}
\renewcommand{\thefigure}{S\arabic{figure}}
\renewcommand{\theequation}{S\arabic{equation}}

\renewcommand\thesection{\Roman{section}}
\renewcommand\thesubsection{\Alph{subsection}}
\setcounter{figure}{0}
\setcounter{equation}{0}

\begin{widetext}
\begin{center}
\textbf{Supplementary Material}    
\end{center}    

	This Supplemental Material provides more details about the inhomogeneous Bloch superconducting (SC) state induced by the parallel magnetic field in two weakly linked SC ultra-thin films, including the discussions on phase transition, fluxoid quantization and an analytical approach to minimize the free energy density. We also briefly discuss the stability of this bilayer SC system in the presence of an extra perpendicular magnetic field.

\tableofcontents

\section{More Details on Generic Bloch Solution}

\subsection{Verification of the gauge invariance of the original model}
In the beginning of the main text, we proposed a two-component Ginzburg-Landau (GL) model Eq.~(1) to describe our bilayer system. For convenience, here is the equation again
\begin{align}
	f[\psi_1(\br),\psi_2(\br)] &= f_n + \sum_{l=1,2}\bigg\{\a|\psi_l(\br)|^2 + \frac{\b}{2}|\psi_l(\br)|^4 
	+ \frac{1}{2m^*}\left|\left(\frac{\hbar}{i}\nabla - \frac{e^*}{c}\bf{A}_l(\br)\right)\psi_l(\br)\right|^2\bigg\} \notag\\
	&\quad + g\left\{\psi_1(\br)^*\psi_2(\br) \exp\bigg[i\frac{2\pi}{\Phi_0}\int_{\bR_2}^{\bR_1}\bf{A}(\bf{s})\cdot\d\bf{s}\bigg] +\m{c.c.}\right\},
\end{align}
where $\br=(x,y)$ denotes a 2D coordinate and $\bR=(x,y,z)$ denotes a 3D one. To be specific, $\bR_l=(x,y,\eta_l\frac{a}{2})$ with short notation $\eta_l$ defined in the main text, and $\bf{A}_l(\br)=\bf{A}(\bR_l)$. This free energy functional should be gauge-invariant under such a transformation
\begin{equation}
	\begin{aligned}
		\psi_l(\br) &\to \psi'_l(\br) = \e^{-i\frac{e^*}{\hbar c}\chi_l(\br)}\psi_l(\br), \\
		\bf{A}(\bR) &\to \bf{A}'(\bR) = \bf{A}(\bR) - \nabla\chi(\bR),
	\end{aligned}
\end{equation}
where $\chi(\bR)$ is an arbitrary scalar function defined in entire space, and $\chi_l(\br)=\chi(\bR_l)$.

Now, let us check the gauge-invariance of our model. For $|\psi_l(\br)|^2$ and $|\psi_l(\br)|^4$ terms, they are obviously invariant under the transformation. For covariant derivative term
\begin{align}
	\frac{1}{2m^*}&\left|\left(\frac{\hbar}{i}\nabla - \frac{e^*}{c}\bf{A}_l(\br)\right)\psi_l(\br)\right|^2 
	\to \frac{1}{2m^*}\left|\left(\frac{\hbar}{i}\nabla - \frac{e^*}{c}\bf{A}'_l(\br)\right)\psi'_l(\br)\right|^2 \notag\\
	&= \frac{1}{2m^*}\left|\left(\frac{\hbar}{i}\nabla - \frac{e^*}{c}\left[\bf{A}_l(\br) - \nabla\chi_l(\br)\right]\right) \left[\e^{-i\frac{e^*}{\hbar c}\chi_l(\br)}\psi_l(\br)\right]\right|^2 \notag\\
	&= \frac{1}{2m^*}\left|\e^{i\frac{e^*}{\hbar c}\chi_l(\br)}\left(\frac{\hbar}{i}\nabla + \frac{\hbar}{i}\left[-\frac{ie^*}{\hbar c}\nabla\chi_l(\br)\right]- \frac{e^*}{c}\bf{A}_l(\br) + \frac{e^*}{c}\nabla\chi_l(\br)\right)\psi_l(\br)\right|^2 \notag\\
	&= \frac{1}{2m^*}\left|\left(\frac{\hbar}{i}\nabla - \frac{e^*}{c}\bf{A}_l(\br)\right)\psi_l(\br)\right|^2,
\end{align}
which is gauge-invariant. And for Josephson coupling term
\begin{align}
	g\psi_1(\br)^*&\psi_2(\br)\exp\bigg[i\frac{2\pi}{\Phi_0}\int_{\bR_2}^{\bR_1}\bf{A}(\bf{s})\cdot\d\bf{s}\bigg] 
	\to g\psi'_1(\br)^*\psi'_2(\br) \exp\bigg[i\frac{2\pi}{\Phi_0}\int_{\bR_2}^{\bR_1}\bf{A}'(\bf{s})\cdot\d\bf{s}\bigg] \notag\\
	&= g\left[\e^{i\frac{e^*}{\hbar c}\chi_1(\br)}\psi_1(\br)^*\right] \left[\e^{-i\frac{e^*}{\hbar c}\chi_2(\br)}\psi_2(\br)\right] \exp\bigg[i\frac{2\pi}{\Phi_0}\int_{\bR_2}^{\bR_1}\left[\bf{A}(\bf{s})-\nabla\chi(\bf{s})\right]\cdot\d\bf{s}\bigg] \notag\\
	&= g\psi_1(\br)^*\psi_2(\br)\,\e^{i\frac{e^*}{\hbar c}[\chi_1(\br)-\chi_2(\br)]}
	\exp\bigg[i\frac{2\pi}{\Phi_0}\int_{\bR_2}^{\bR_1}\bf{A}(\bf{s})\cdot\d\bf{s}\bigg] \e^{-i\frac{2\pi}{\Phi_0}[\chi(\bR_1)-\chi(\bR_2)]} \notag\\
	&= g\psi_1(\br)^*\psi_2(\br)\exp\bigg[i\frac{2\pi}{\Phi_0}\int_{\bR_2}^{\bR_1}\bf{A}(\bf{s})\cdot\d\bf{s}\bigg],
\end{align}
where we used $\Phi_0=hc/e^*$ and $\chi_l(\br)=\chi(\bR_l)$ in the last equal sign. Therefore our model Eq.~(1) is gauge-invariant.

In the main text, we chose the Landau gauge $\bf{A}=Hz\ex$, and we found that it does not appear in the Josephson coupling term since $\int_{\m{R}_2}^{\m{R}_1}\bf{A}\cdot\d\bf{s}=0$. In fact, we could choose another gauge for vector potential, say $\bf{A}=-Hx\ez$. The ``$\nabla$" in covariant derivative term is actually a 2D gradient such that the $z$-component of $\bf{A}$ does not contribute to the kinetic part, but the gauge-invariant phase of the coupling term, namely
\begin{equation*}
	\frac{2\pi}{\Phi_0}\int_{\bR_2}^{\bR_1}\bf{A}(\bf{s})\cdot\d\bf{s} = \frac{2\pi}{\Phi_0}\int_{-a/2}^{a/2}\left(-Hx\right)\d z = -\frac{2\pi Hx}{\Phi_0}a = -k_0x,
\end{equation*}
where $k_0=2\pi Ha/\Phi_0$. Denote the SC order parameter $\tpsi_l(\br)$ in this new gauge, the GL free energy functional in this gauge reads
\begin{align}\label{gauge-inv}
	f[\tpsi_1(\br),\tpsi_2(\br)] &= f_n + \sum_{l=1,2}\bigg\{\a|\tpsi_l(\br)|^2 + \frac{\b}{2}|\tpsi_l(\br)|^4 
	+ \frac{\hbar^2}{2m^*}\left|\nabla\tpsi_l(\br)\right|^2\bigg\} \notag\\
	&\quad + g\left[\tpsi_1(\br)^*\tpsi_2(\br)\,\e^{-ik_0x} + \tpsi_2(\br)^*\tpsi_1(\br)\,\e^{ik_0x}\right],
\end{align}
which is exactly the free energy functional given in gauge $\bf{A}=Hz\ex$ (Eq.~(2) in the main text) re-expressed after a gauge transformation $\psi_l(\br) = \e^{i\eta_l(k_0/2)x}\tpsi_l(\br)$. After reduced to 1D model and by brief discussion on periodicity, this would exactly give a Bloch solution of SC order parameter.

Note that the FF state $\tpsi_l^{\m{FF}}(\br) = \sqrt{\rho_{\m{s}}^{\m{FF}}}\e^{-i\eta_l(k_0/2)x}\e^{i\vp_l}$ is an exact solution to Eq.~\eqref{gauge-inv} in the aforementioned gauge $\bf{A}=-Hx\ez$, the physical properties of this solution are consistent with the discussions in the main text where we chose the gauge $\bf{A}=Hz\ex$.

\subsection{Discrete Ginzburg-Landau equations and solution structure of the system}
Starting from the reduced 1D GL free energy functional, Eq.~(2) in the main text,
the main difficulty in seeking a generic solution that minimizes the free energy is to treat the covariant derivative term $(\hbar^2/2m^*)|(\pt_x-i\eta_{l}k_0/2)\psi_{l}|^2$. Inspired by Ref.~\cite{Pethick}, through analysis we confine our attention to solutions of a Bloch form $\psi_l(x) = \e^{i\eta_l(k_0/2)x}\tpsi_l(x)$ with $\tpsi_l(x+a_0)=\tpsi_l(x)$ and $a_0=2\pi/k_0$. Substituting it into the GL free energy functional Eq.~(2) yields
\begin{equation}\label{GLfe}
	f[\tpsi_1,\tpsi_2] = f_n + \sum_{l=1,2}\bigg(\a|\tpsi_l|^2 + \frac{\b}{2}|\tpsi_l|^4 + \frac{\hbar^2}{2m^*}\bigg|\pp{\tpsi_l}{x}\bigg|^2\bigg) + g\left(\e^{-ik_0x}\tpsi_1^*\tpsi_2 + \e^{+ik_0x}\tpsi_2^*\tpsi_1\right).
\end{equation}
Apparently, it has a period $a_0$, namely $f[\tpsi_1(x+a_0),\tpsi_2(x+a_0)] = f[\tpsi_1(x),\tpsi_2(x)]$. So far we have managed to simplify the covariant derivative terms which would give rise to large complexity in calculation, but the cost is that Josephson coupling strength becomes periodically modulated. Thanks to periodicity, we could expand $\tpsi_l$ into plane waves
\begin{equation}
	\tilde{\psi}_l(x) = \sqrt{n}\sum_{\nu=-\infty}^{\infty}a_{l\nu}\e^{i\nu k_0x}.
\end{equation}
where $\nu$ is an integer, and
\begin{equation}\label{avgsfden}
	n = \frac{1}{a_0}\int_{0}^{a_0}|\tilde{\psi}_l(x)|^2\d x
\end{equation}
is the average superfluid density per period for each layer. Hence the coefficients follows normalization relation
\begin{equation}\label{condition}
	\sum_{\nu=-\infty}^{\infty}|a_{l\nu}|^2 = 1.
\end{equation}

A standard procedure to find the generic solution is to first calculate the average free energy per unit volume, or free energy density, then we search its minimum by optimizing the free energy density about variables at all degrees of freedoms, at some set of given physical parameters. The free energy density is defined as
\begin{equation}
	\fed \equiv \frac{1}{a_0}\int_{0}^{a_0}\d x\left(f[\tilde{\psi}_1(x),\tilde{\psi}_2(x)]-f_n\right).
\end{equation}
After integration, the free energy density is re-expressed in terms of $\{a_{l\nu}\}$ and $n$, namely
\begin{align}\label{bloch}
	\fed &= \sum_{l=1,2}\bigg(n\a\sum_{\nu}|a_{l\nu}|^2 + \frac{n^2\b}{2}\sum_{\nu_1,\nu_2,\nu_3} a_{l\nu_1}^*a_{l\nu_2}^*a_{l\nu_3}a_{l,\nu_1+\nu_2-\nu_3} + n\frac{\hbar^2k_0^2}{2m^*}\sum_{\nu}\nu^2|a_{l\nu}|^2 + ng\sum_{\nu}a_{l\nu}^*a_{\bar{l},\nu+\eta_l}\bigg) \notag\\
	&= n\sum_{l=1,2}\sum_{\nu}\left(\frac{\hbar^2k_0^2}{2m^*}\nu^2 +\a\right)|a_{l\nu}|^2 + \frac{n^2\b}{2}\sum_{l=1,2}\sum_{\nu_1,\nu_2,\nu_3} a_{l\nu_1}^*a_{l\nu_2}^*a_{l\nu_3}a_{l,\nu_1+\nu_2-\nu_3}\notag\\
	&\quad + ng\sum_{\nu}\left(a_{1\nu}^*a_{2,\nu+1}+a_{2\nu}^*a_{1,\nu-1}\right).
\end{align}
Taking variations with respect to $\{a_{l\nu}^*\}$ and $n$ yields the discrete GL equations and self-consistent equation Eq.~(9) in the main text. In our practical study, we optimize the free energy density $\fed$ by introduce Lagrangian multipliers
\begin{equation*}
	\mathcal{L.M.} = n\sum_{l=1,2}\l_l\bigg(\sum_{\nu}|a_{l\nu}|^2-1\bigg)
\end{equation*}
and solve the equations $\pt(\fed+\mathcal{L.M.})/\pt{a_{l\nu}^*}=0$ and $\pt(\fed+\mathcal{L.M.})/\pt{n}=0$ numerically, such that the obtained solution set $\{a_{l\nu}\}$ spontaneously satisfies the normalization condition Eq.~\eqref{condition}.

As discussed in the main text, it is more convenient to solve $\{n\beta,a_{1\nu},a_{2\nu}\}$ instead of $\{n,a_{1\nu},a_{2\nu}\}$, such that the parameter $\beta$ will be irrelevant to the solutions. The optimized free energy density $\fed_\m{Bloch}\propto n \propto \b^{-1}$ since the solution $\{n\beta,a_{1\nu},a_{2\nu}\}$ is a set of definite numbers for a given set of physical parameters $(|g|/|\a|,H)$. While in previous discussion we also find that $\fed_\m{FF} \propto \b^{-1}$. Hence the phase diagram for the ground state (Fig.~2 in the main text) is indeed independent of $\b$. In order to show the solution structure in the phase diagram more clearly, we plot both the free energy densities $\fed_\m{Bloch}$ and $\fed_\m{FF}$ versus $H$ at fixed $|g|/|\a|$, shown in Fig.~\ref{FHmap}. 

\begin{figure}[htbp]
	\centering
	\includegraphics[width=\linewidth]{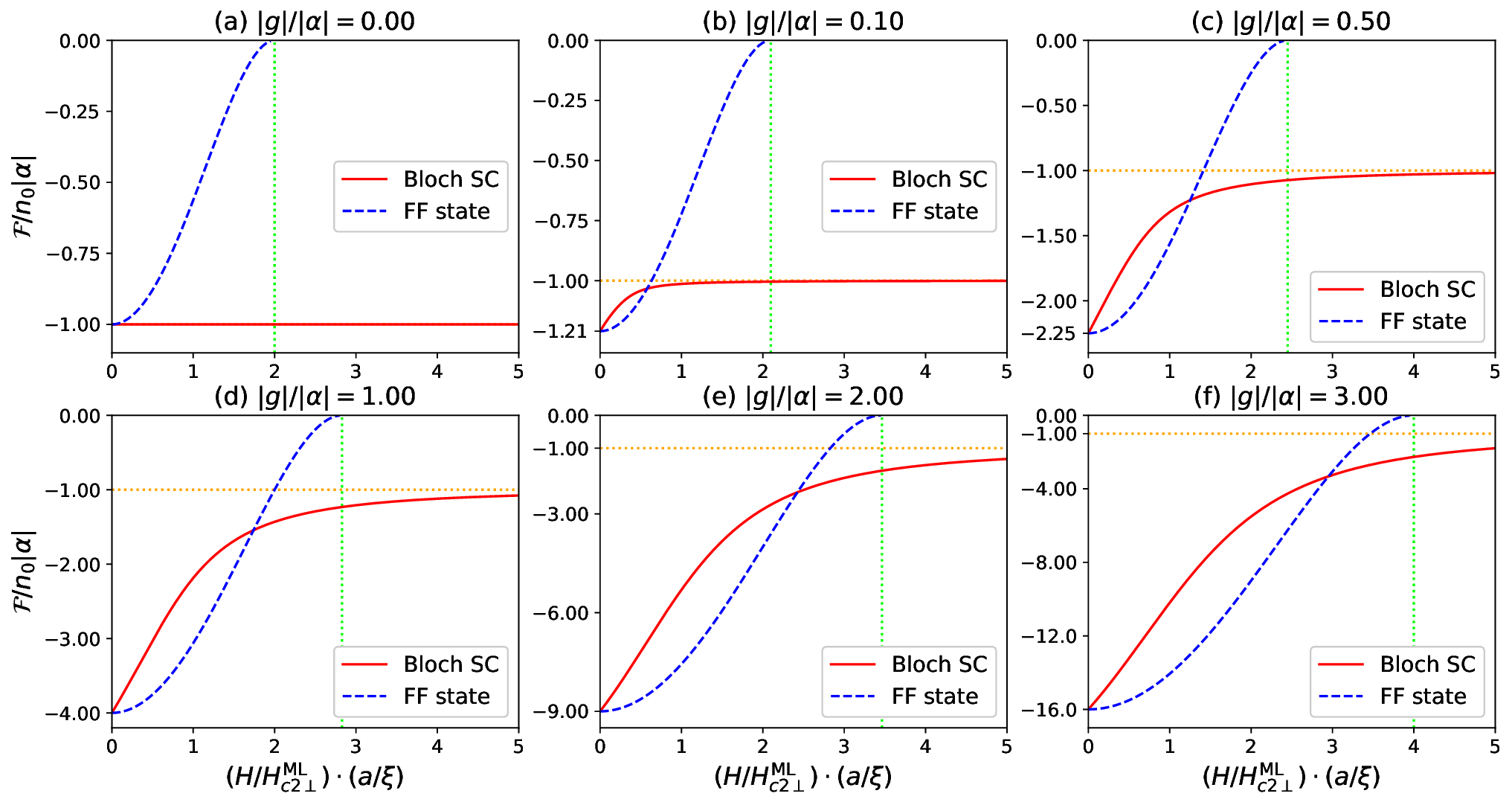}
	\caption{Free energy density $\fed$ of Bloch SC (red solid curve) and FF state (blue dashed curve) solutions versus magnetic field $H$ at fixed $|g|/|\a|$. We measure $\fed$ in units $n_0|\a|$, where $n_0=|\a|/\b$. FF state solution exists up to $H_\m{c}^*$ (green dotted vertical line). Subplot (a) indicates that at $g=0$, the Bloch SC solution is exactly an decoupled SC state, with $\fed_\m{Bloch}|_{g=0}$ fixed at $\fed_\m{D}=-\a^2/\b=-n_0|\a|$ regardless of magnetic field. At $H=0$ the Bloch solution restores to FF state. At finite $g$ and $H$, the lower one of the two curves represents the ground state free energy density of the system. Besides, we find $\fed_\m{Bloch}\to\fed_\m{D}$ (orange dotted horizontal line) as $H\to\infty$.}
	\label{FHmap}
\end{figure}

Here we measure the free energy density in unit $n_0|\a|$, where $n_0=|\a|/\b$ is the superfluid density for a single layer without magnetic field. In Fig.~\ref{FHmap}, the red solid curve represents $\fed_\m{Bloch}$ which is obtained from numerical study, whereas the blue dashed curve for $\fed_\m{FF}$ has an explicit dimensionless expression in the plot
\begin{equation*}
	\frac{\fed_\m{FF}}{n_0|\a|} = -\left[1-\frac{1}{4}\left(\frac{Ha}{H_{\m{c}2\perp}^{\mathrm{ML}}\xi}\right)^2+\frac{|g|}{|\a|}\right]^2.
\end{equation*}

These plots tell us at least three pieces of information. 

(1) At finite $H$ and $g$, the ground state changes from FF state to Bloch SC state as $H$ increases at given $g$. As $H\to\infty$, the free energy density $\fed_\m{Bloch}\to-n_0|\a|=-\a^2/\b$, with the average superfluid density Eq.~\eqref{avgsfden} $n_\m{Bloch}\to n_0$, meanwhile for decoupled SC state, these values are $\fed_\m{D}=-\a^2/\b$ and $n_\m{D}=\rho_{\m{s}0}=n_0=|\a|/\b$, confirming relevant discussion in the main text. (Note $n_\m{D}$ is average superfluid density per period and $\rho_{\m{s}0}$ is local superfluid density.) The FF state solution could exist up to $H_\m{c}^*$ (the green dotted vertical line) while the transition occurs at some value $H$ lower than $H_\m{c}^*$, which explains why the intersection of solid green and yellow zone is below the green dashed line for $H_\m{c}^*$ in the ground state phase diagram Fig.~2 in the main text. For small field $H$, Bloch SC solution exists but possesses a higher free energy than FF state.

(2) Fig.~\ref{FHmap}(a) indicates that when $g=0$, the free energy density of Bloch SC state is unchanged at any $H$, whose value exactly equals to $\fed_\m{Bloch}|_{g=0}=\fed_\m{D}=-\a^2/\b$. Numerical result tells us that $a_{l\nu}=\de_{\nu0}$, as well as $n_\m{Bloch}|_{g=0}=n_\m{D}=|\a|/\b$, indicating that the Bloch solution is exactly an decoupled SC state at $g=0$, in agreement with Eq.~(6) in the main text. 

(3) At $H=0$ the two curves starts from the same point on the vertical axis, which illustrates that without external magnetic field, ``both" solutions have the same free energy density. However, $H=0$ means $k_0=0$, so the ``transformed" free energy density Eq.~\eqref{GLfe} restore to the original one (Eq.~(2) in the main text) with $k_0=0$. Hence at zero magnetic field, there is only one FF state solution to the GL equation, $\psi_l(x)|_{H=0}= \sqrt{\rho_\m{s}^\m{FF}|_{H=0}}$, where $\rho_\m{s}^\m{FF}|_{H=0} = -(\a-|g|)/\b$. The free energy density at $H=0$ is $\fed_\m{FF}|_{H=0}=-(\a-|g|)^2/\b=-n_0|\a|(1+|g|/|\a|)^2$.

\subsection{Commensurate-Incommensurate phase transition}
According to phase diagram Fig.~2 in the main text and Fig.~\ref{FHmap}, there is a phase transition from FF state to Bloch SC state as magnetic field $H$ increases at fixed tunneling strength $g$ (or as $g$ decreases at fixed $H$). This phase transition is a commensurate-incommensurate (CI) transition and is of second order. 

To see this, we start with the FF state to derive the effective theory for the phase transition. In a superconductor, it is well known that the fluctuations of superfluid density (or the amplitude of the SC order parameter) is massive due to the Higgs mechanism. For a low energy theory, it is sufficient to treat it as a constant $\rho_{\m{s}}$ and assume that the fluctuations come from the local phase, or actually what matters is the relative phase $\phi(\br)$ between two layers. In the absence of external magnetic field, the effective Hamiltonian takes the form
\begin{equation}
	H_\m{eff} = \rho_{\m{s}} \int\d^2\br\,\frac{\hbar^2}{2m^*}|\nabla\phi(\br)|^2 + H_\m{T}.
\end{equation}
By choosing the gauge $\bf{A}=-Hx\ez$, the tunneling energy term reads
\begin{align*}
	H_\m{T} &= \int\d^2\br\,g\left(\psi_1^*(\br)\psi_2(\br)\e^{-ik_0x} + \psi_2^*(\br)\psi_1(\br)\e^{ik_0x}\right) \\
	&= \int\d^2\br\, 2g|\psi_1(\br)||\psi_2(\br)| \cos\left[\phi_2(\br)-\phi_1(\br)-k_0x\right] \\
	&\simeq \rho_{\m{s}} \int\d^2\br\, 2g\cos\left[\phi(\br)-k_0x\right].
\end{align*}
Hence the effective Hamiltonian becomes
\begin{equation}
	H_\m{eff} = \rho_{\m{s}} \int\d^2\br\,\bigg\{\frac{\hbar^2}{2m^*}|\nabla\phi(\br)|^2 + 2g\cos\left[\phi(\br)-k_0x\right]\bigg\},
\end{equation}
which is precisely the Pokrovsky-Talapov (PT) model~\cite{KYang94,KYang96}, describing a CI transition~\cite{PT79}. 

For small $k_0$ or small $H$ field, the FF state has lower free energy, which is given by $\nabla\phi(\br)$ and $\phi(\br)=k_0x$ under the gauge choice $\bf{A}=-Hx\ez$ and corresponds to a commensurate state. As $H$ increases, the system undergoes a continuous phase transition to an incommensurate state, namely, the Bloch SC state obtained by our GL mean-field theory, which breaks the translational symmetry.

By the way, in another gauge choice $\bf{A}=Hz\ex$, which is adopted for most of the numerical and analytical calculations in our main study, we have
\begin{equation}
	H_\m{eff} = \rho_{\m{s}} \int\d^2\br\,\bigg\{\frac{\hbar^2}{2m^*}\left(|\pt_x\vp(\br)+k_0|^2 + |\pt_y\vp(\br)|^2\right) + 2g\cos\vp(\br)\bigg\}.
\end{equation}
Similar discussion follows.

\subsection{Spatial modulation of the order parameters and fluxoid quantization}
We are also interested in the spatial modulation of the order parameters $\psi_l(x)$, by definition
\begin{equation}\label{blochop}
	\psi_l(x) = \e^{i\eta_l\frac{k_0}{2}x}\tpsi_l(x) = \sqrt{n}\,\e^{i\eta_l\frac{k_0}{2}x} \sum_{\nu=-\infty}^{\infty}a_{l\nu}\e^{i\nu k_0x} = \sqrt{n}\sum_{\nu=-\infty}^{\infty}a_{l\nu}\e^{i\left(\nu+\frac{\eta_l}{2}\right)k_0x}.
\end{equation}
As discussed in the main text, the Bloch solution obeys the $\cal{T}\cal{P}_{z}$ symmetry such that $\psi_{\bar{l}}(x)^{*}=\psi_{l}(x)$, which means
\begin{align*}
	\psi_{\bar{l}}(x)^* - \psi_{l}(x) &= \sqrt{n}\sum_{\nu=-\infty}^{\infty}a^*_{\bar{l}\nu}\e^{-i\left(\nu+\frac{\eta_{\bar{l}}}{2}\right)k_0x} - \sqrt{n}\sum_{\nu=-\infty}^{\infty}a_{l\nu}\e^{i\left(\nu+\frac{\eta_l}{2}\right)k_0x} \\
	&= \sqrt{n}\sum_{\nu=-\infty}^{\infty}\left(a^*_{\bar{l}\nu} - a_{l,-\nu}\right) \e^{-i\left(\nu-\frac{\eta_{l}}{2}\right)k_0x} \equiv 0,
\end{align*}
where we used $\eta_{\bar{l}}=-\eta_{l}$ by definition of this notation. This indicates that
\begin{equation}\label{symcond}
	a^*_{\bar{l}\nu}=a_{l,-\nu}.
\end{equation}

Besides, we mentioned that if $\{n\beta,a_{1\nu},a_{2\nu}\}$ is a solution to Eqs.~(9) for a given pair of $(g/|\alpha|,H)$, then $\{n\beta,e^{i\pi\nu}a_{1\nu},e^{i\pi\nu}a_{2\nu}\}$ will be a solution for $(-g/|\alpha|,H)$. Here below is the proof. First, let us check whether this set of solution satisfies the symmetry condition Eq.~\eqref{symcond}. If $g\to g'=-g$, denote 
\begin{equation}\label{gtomg}
	a_{l\nu}\to a'_{l\nu}=\e^{i\pi\nu}a_{l\nu},
\end{equation}
and indeed $a'^*_{\bar{l}\nu} = \e^{-i\pi\nu}a_{\bar{l}\nu}^* = \e^{i\pi(-\nu)}a_{l,-\nu} = a'_{l,-\nu}$. Second, we demonstrate that the free energy density Eq.~\eqref{bloch} is unchanged after this ``transformation". The first quadratic term has nothing to do with phases of $\{a_{l\nu}\}$, the phases cancel in the second quartic term, while for the third term, or coupling term
\begin{equation*}
	V = ng\sum_{\nu}\left(a_{1\nu}^*a_{2,\nu+1}+a_{2\nu}^*a_{1,\nu-1}\right),
\end{equation*} 
we have
\begin{align*}
	V \to V' 
	&= ng'\sum_{\nu}\left(a'^*_{1\nu}a'_{2,\nu+1}+a'^*_{2\nu}a'_{1,\nu-1}\right) 
	= -ng\sum_{\nu}\left(\e^{-i\pi\nu}a^*_{1\nu}\e^{i\pi(\nu+1)}a_{2,\nu+1} +\e^{-i\pi\nu}a^*_{2\nu}\e^{i\pi(\nu-1)}a_{1,\nu-1}\right) \\
	&= -ng\sum_{\nu}\left(\e^{i\pi}a^*_{1\nu}a_{2,\nu+1}+\e^{-i\pi}a^*_{2\nu}a_{1,\nu-1}\right) 
	= ng\sum_{\nu}\left(a^*_{1,\nu-1}a_{2,\nu}+a_{1,\nu}a^*_{2,\nu+1}\right) = V.
\end{align*}
Therefore the physics is independent on the sign of $g$.

In our numerical study, the truncation for the series $\{a_{1\nu}\}$ and $\{a_{2\nu}\}$ has to been introduced, namely $a_{1\nu}=a_{2\nu}=0$ for $|\nu|>\nu_{\max}$, where $\nu_{\max}$ is a positive integer. We use iterations to realize this truncation. In the $t$-th step, we set $\nu_{\max}=t$, solving the nonlinear equations we could obtain a set of $\{a_{l\nu}^{(t)}\}$ as well as free energy density $\fed^{(t)}$. We truncate the iteration process when the free energy density converges at some $t$, where the error of free energy density with respect to the last step is smaller than some preset tolerance $\eps$, namely $|(\fed^{(t)}-\fed^{(t-1)})/\fed^{(t-1)}|<\eps$. We set $\eps=0.00005$ in our practical study. 

\begin{minipage}{\textwidth}
	\begin{minipage}{\textwidth}
		\centering
		\makeatletter\def\@captype{table}
		\caption{Numerical result of $\{a_{l\nu}\}$ for (A) state with $g>0$.}
		\label{positiveg}
		\begin{tabular}{p{1cm}|rrrrrrr}
			\hline
			\hline
			\makecell[c]{$a_{l\nu}$} & 
			\makecell[c]{$\nu=-3$} & 
			\makecell[c]{$\nu=-2$} & 
			\makecell[c]{$\nu=-1$} &
			\makecell[c]{$\nu=0$} &
			\makecell[c]{$\nu=1$} & 
			\makecell[c]{$\nu=2$} & 
			\makecell[c]{$\nu=3$} \\ 
			\hline
			\makecell[c]{$l=1$} & 
			$-2.33\times10^{-4}$ &
			$-6.22\times10^{-3}$ &
			$-2.27\times10^{-1}$ &
			$9.72\times10^{-1}$ &
			$5.64\times10^{-2}$ &
			$2.31\times10^{-3}$ &
			$9.32\times10^{-5}$ \\ 
			\makecell[c]{$l=2$} &
			$9.32\times10^{-5}$ &
			$2.31\times10^{-3}$ &
			$5.64\times10^{-2}$ &
			$9.72\times10^{-1}$ &
			$-2.27\times10^{-1}$ &
			$-6.22\times10^{-3}$ &
			$-2.33\times10^{-4}$ \\
			\hline
			\hline
		\end{tabular}
	\end{minipage}
	\begin{minipage}{\textwidth}
		\centering
		\makeatletter\def\@captype{table}
		\caption{Numerical result of $\{a_{l\nu}\}$ for (A) state with $g<0$.}
		\label{negativeg}
		\begin{tabular}{p{1cm}|rrrrrrr}
			\hline
			\hline
			\makecell[c]{$a_{l\nu}$} & 
			\makecell[c]{$\nu=-3$} & 
			\makecell[c]{$\nu=-2$} & 
			\makecell[c]{$\nu=-1$} &
			\makecell[c]{$\nu=0$} &
			\makecell[c]{$\nu=1$} & 
			\makecell[c]{$\nu=2$} & 
			\makecell[c]{$\nu=3$} \\ 
			\hline
			\makecell[c]{$l=1$} & 
			$2.33\times10^{-4}$ &
			$-6.22\times10^{-3}$ &
			$2.27\times10^{-1}$ &
			$9.72\times10^{-1}$ &
			$-5.64\times10^{-2}$ &
			$2.31\times10^{-3}$ &
			$-9.32\times10^{-5}$ \\ 
			\makecell[c]{$l=2$} &
			$-9.32\times10^{-5}$ &
			$2.31\times10^{-3}$ &
			$-5.64\times10^{-2}$ &
			$9.72\times10^{-1}$ &
			$2.27\times10^{-1}$ &
			$-6.22\times10^{-3}$ &
			$2.33\times10^{-4}$ \\
			\hline
			\hline
		\end{tabular}	
	\end{minipage}
	\vspace{10pt}
\end{minipage}

The numerical results for (A) state marked by the red star in Fig.~2 with both positive and negative $g$ are listed in Tab.~\ref{positiveg} and Tab.~\ref{negativeg} respectively. Obviously, the solutions satisfy symmetry condition Eq.~\eqref{symcond}, as well as phase condition Eq.~\eqref{gtomg} for different sign of $g$. Since the order parameters are complex-valued functions, we plot the spatial modulation of their real and imaginary parts in Fig.~\ref{op} according to numerical calculation. The order parameters have period $2a_0$ while that for the local superfluid density $|\psi_l|^2$ is $a_0$, where $a_0=2\pi/k_0=\Phi_0/Ha$. The parameters are chosen at two red stars (A and B states) in Fig.~2 for Figs. \ref{op18} and \ref{op40}. As the in-plane magnetic field $H$ increases, both real and imaginary parts gradually change to appear like trigonometric functions. The spatial modulation of superfluid density becomes not so distinguishable. At a large enough magnetic field (if it does not exceed the Pauli limit $H_\m{P}$), the superfluid density is almost constant, shown in Fig.~\ref{op200}.

\begin{figure}[htbp]
	\centering
	\subfigure{
		\label{op18}
		\begin{overpic}[width=\linewidth]{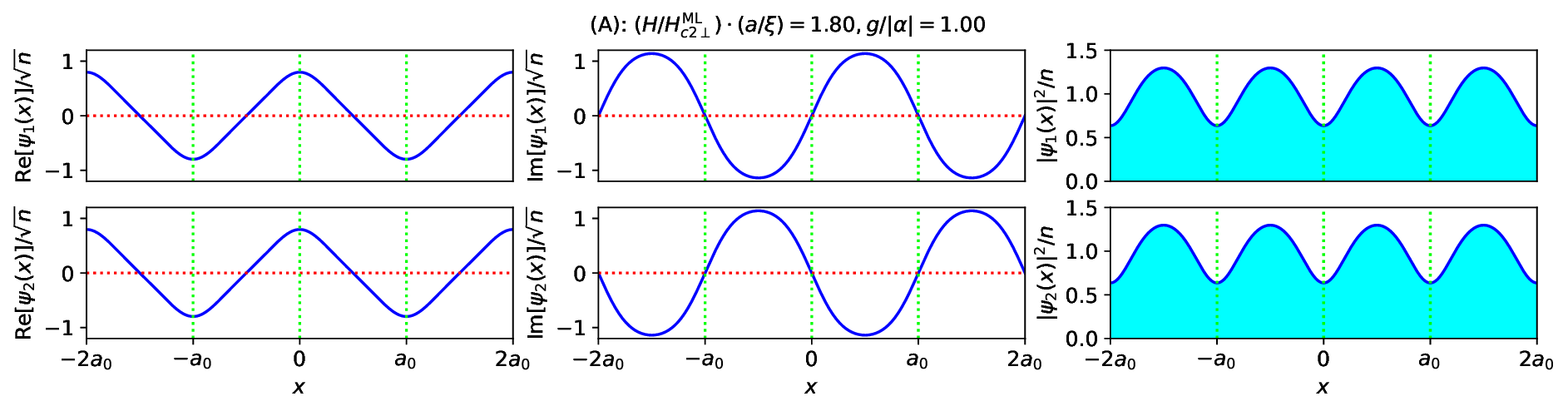}
			\put(0,24){(a)}
		\end{overpic}
	}
	\vfill
	\subfigure{
		\label{op40}
		\begin{overpic}[width=\linewidth]{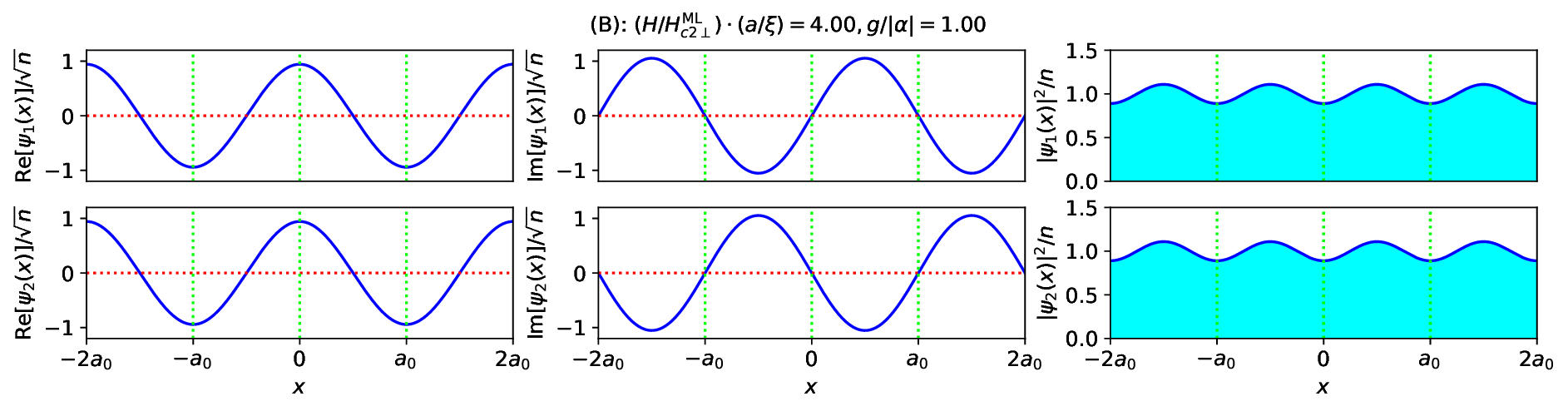}
			\put(0,24){(b)}
		\end{overpic}
	}
	\vfill
	\subfigure{
		\label{op200}
		\begin{overpic}[width=\linewidth]{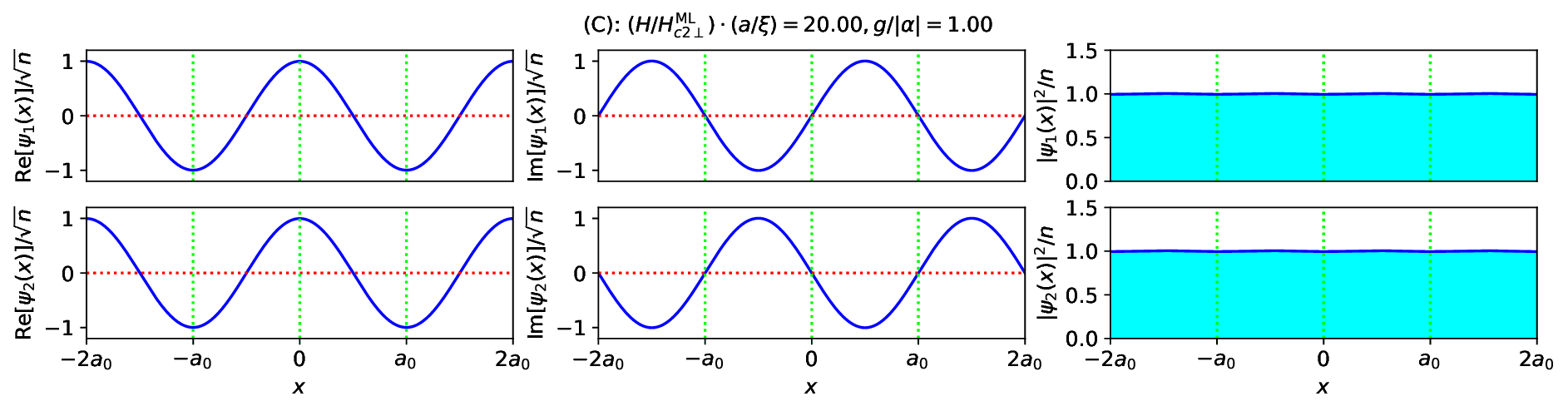}
			\put(0,24){(c)}
		\end{overpic}
	}
	\caption{Order parameters and superfluid density for both layers in gauge $\bf{A}=Hz\ex$. (a) Bloch SC state (A) in Fig.~2. (b) Bloch SC state (B) in Fig.~2. (c) Bloch SC state at a large enough magnetic field.}
	\label{op}
\end{figure}

Obtaining the spatial modulation of order parameters $\psi_l(x)$ for generic Bloch SC solution, we could calculate supercurrent density and prove that the fluxoid per period $\Phi'$ is quantized and equals to $\Phi_0$. Consider a loop enclosing a period of $a_0$ (indicated by green dashed loops in Fig.~3 in main text), the London's fluxoid is defined as
\begin{equation}
	\Phi' = \Phi + \frac{m^{*}c}{e^{*2}}\oint\frac{\bf{J}_{\m{s}}}{\rho_{\m{s}}}\cdot\d\bf{s},
\end{equation}
where $\Phi=\oint\bf{A}\cdot\d\bf{s}=Haa_0=\Phi_0$ is the the ordinary flux, and for a Bloch SC state, we have $J_{\m{T}}(x+a_0)=J_{\m{T}}(x)$, which does not contribute to flux. The only contribution to the second part in the fluxoid comes from
\begin{equation}\label{Jsdef}
	\bf{J}_{\m{s}l} = \frac{e^*\hbar}{2m^*i}\left(\psi_l^*\nabla\psi_l-\psi_l\nabla\psi_l^*\right) - \frac{e^{*2}}{m^*c}\psi_l^*\psi_l\bf{A}_l.
\end{equation}
We could re-express the spatially modulated complex order parameter as
\begin{equation}
	\psi_l(x) = \sqrt{\rho_{\m{s}l}(x)}\,\e^{i\vp_l(x)},
\end{equation}
where $\rho_{\m{s}}(x)=|\psi_l(x)|^2$ is defined as local superfluid density. And
\begin{align*}
	\pt_x\psi_l &= \left[\left(\pt_x\sqrt{\rho_{\m{s}l}}\right)+i\sqrt{\rho_{\m{s}l}}\left(\pt_x\vp_l\right)\right]\e^{i\vp_l}, \\
	\psi_l^*\pt_x\psi_l &= \sqrt{\rho_{\m{s}l}}\left(\pt_x\sqrt{\rho_{\m{s}l}}\right)+i\rho_{\m{s}l}\left(\pt_x\vp_l\right).
\end{align*}
The supercurrent $\bf{J}_{\m{s}l}$ only has $x$ component, which is
\begin{equation*}
	J_{\m{s}l}(x) = \frac{e^*\hbar}{2m^*i}\left(\psi_l^*\pt_x\psi_l-\psi_l\pt_x\psi_l^*\right) - \frac{e^{*2}}{m^*c}|\psi_l|^2A_l = \left[\frac{e^*\hbar}{m^*}\pt_x\vp_l(x) - \frac{e^{*2}Ha}{2m^*c}\eta_{l}\right]\rho_{\m{s}l}(x).
\end{equation*}
and the second part in the fluxoid reads
\begin{align*}
	\frac{m^{*}c}{e^{*2}}\oint\frac{\bf{J}_{\m{s}}(x)}{\rho_{\m{s}}(x)}\cdot\d\bf{s} &= \frac{m^{*}c}{e^{*2}}\oint\left[\frac{e^*\hbar}{m^*}\pt_x\vp_l(x) - \frac{e^{*2}Ha}{2m^*c}\eta_{l}\right]\ex\cdot\d\bf{s} 
	= \frac{\hbar c}{e^*}\oint \pt_x\vp_l(x)\,\ex\cdot\d\bf{s} - \frac{Ha}{2}\oint \eta_{l}\,\ex\cdot\d\bf{s} \\
	&= \frac{\hbar c}{e^*}\left[\int_{0}^{a_0} \pt_x\vp_1(x)\, \d x + \int_{a_0}^{0} \pt_x\vp_2(x)\, \d x\right] - \frac{Ha}{2}\left[\int_{0}^{a_0} \left(+1\right)\d x + \int_{a_0}^{0} \left(-1\right)\d x \right] \\
	&= \frac{\Phi_0}{2\pi}\int_{0}^{a_0} \pt_x\left[\vp_1(x)-\vp_2(x)\right] \d x - Haa_0
	= \frac{\Phi_0}{2\pi}\int_{0}^{a_0} \pt_x\left[\De\vp(x)\right] \d x - \Phi_0.
\end{align*}
As shown in Fig.~3, numerical result tells us that the phase difference $\De\vp(x)$ undergoes a $2\pi$ jump at $x=a_0/2$, the integral in the last line must be performed as follows
\begin{align*}
	\frac{\Phi_0}{2\pi}\int_{0}^{a_0} \pt_x\left[\De\vp(x)\right] \d x 
	&= \frac{\Phi_0}{2\pi}\int_{0}^{\frac{a_0}{2}-0^+} \pt_x\left[\De\vp(x)\right] \d x 
	+ \frac{\Phi_0}{2\pi}\int_{\frac{a_0}{2}+0^+}^{a_0} \pt_x\left[\De\vp(x)\right] \d x \\
	&= \frac{\Phi_0}{2\pi}\left[\De\vp\left(\frac{a_0}{2}-0^+\right)-\De\vp(0)\right] + \frac{\Phi_0}{2\pi}\left[\De\vp(a_0)-\De\vp\left(\frac{a_0}{2}+0^+\right)\right] \\
	&= \frac{\Phi_0}{2\pi}(\pi+\pi) = \Phi_0.
\end{align*}
This derivation tells us that the net orbital magnetization is counted by the flux
\begin{equation}\label{netmagzero}
	\frac{m^{*}c}{e^{*2}}\oint\frac{\bf{J}_{\m{s}}}{\rho_{\m{s}}}\cdot\d\bf{s} = 0
\end{equation}
and vanished. Therefore, for Bloch SC state the London's fluxoid reads $\Phi'_\m{Bloch}=\Phi_0$ in each period $a_0$, which is exact the flux quantum. 

For FF state, the derivation of fluxoid quantization is quite simple and yields $\Phi'_{\m{FF}}=0$, as mentioned in the main text. In contrast to Bloch SC state, the FF state manifests perfect diamagnetism.

\subsection{Approximate analytical solutions}
As mentioned in the main text, the Bloch SC state can be well approximated by Jacobian elliptic functions. Here below we will keep our notations simple until we make comparisons with numerical results. Analytically we employ the following trial functions as approximate order parameters
\begin{equation}\label{approx}
	\psi_l(x) = \left[\mu\cn(ux;r)+i\eta_l\nu\sn(ux;r)\right]\e^{i\vp_l},
\end{equation}
where $\mu$, $\nu$ and $r$ are real constants and treated as variational parameters. The overall phase factors are set as $\vp_1=\vp_2=0$ for $g>0$ and $\vp_1=-\vp_2=-\pi/2$ for $g<0$, which have no effect on physics. These order parameters have a spatial period $4K(r)/u$, where $K(r)$ is complete elliptic integrals of the first kind and $r\in(0,1)$ is called the modulus of elliptic integrals. After doing the math we find that if we fix $4K(r)/u=2a_0$ to determine $u$, in agreement with numerical solution (Fig.~\ref{op}), the free energy functional $f[\psi_1(x),\psi_2(x)]$ given by Eq.~(2) in the main text has a spatial period $a_0$, which is well consistent with all previous discussions based on numerical study. The free energy density is calculated in terms of our trial functions
\begin{equation}
	\fed = \frac{1}{a_0}\int_{0}^{a_0}\d x\left(f[\psi_1(x),\psi_2(x)]-f_n\right).
\end{equation}
Through careful derivation, we arrive at a multi-variable function of $\mu$, $\nu$ and $r$, which takes the form 
\begin{align}\label{anafe}
	\fed(\mu,\nu;r) = A(r)\mu^4 + B(r)\mu^2\nu^2 + C(r)\nu^4 + D(r)\mu^2 + F(r)\nu^2 + G(r)\mu\nu.
\end{align}
The approximate free energy density is a quartic function of two variables $\mu$ and $\nu$, the coefficients of each term are only functions of the elliptic modulus $r$. To be specific, they are
\begin{align*}
	A(r) &= \b\frac{(r-1)(3r-2)}{3r^2} + 2\b\left(\frac{2r-1}{3r^2}\right)\frac{E(r)}{K(r)}, \\
	B(r) &= 4\b\left(\frac{r-1}{3r^2}\right) + 2\b\left(\frac{2-r}{3r^2}\right)\frac{E(r)}{K(r)}, \\
	C(r) &= \b\left(\frac{2+r}{3r^2}\right) - 2\b\left(\frac{1+r}{3r^2}\right)\frac{E(r)}{K(r)}, \\
	D(r) &= 2\bigg[\bigg(\a+\frac{\hbar^2k_0^2}{8m^*}+|g|\bigg)\frac{r-1}{r} +\frac{\hbar^2k_0^2}{2m^*}\bigg(\frac{K(r)}{\pi}\bigg)^2 \bigg(\frac{1-r}{3r}\bigg)\bigg] \\
	&\quad +2\bigg[\bigg(\a+\frac{\hbar^2k_0^2}{8m^*}+|g|\bigg)\frac{1}{r}  +\frac{\hbar^2k_0^2}{2m^*}\bigg(\frac{K(r)}{\pi}\bigg)^2 \bigg(\frac{2r-1}{3r}\bigg)\bigg]\frac{E(r)}{K(r)}, \\
	F(r) &= 2\bigg[\bigg(\a+\frac{\hbar^2k_0^2}{8m^*}-|g|\bigg)\frac{1}{r}  +\frac{\hbar^2k_0^2}{2m^*}\bigg(\frac{K(r)}{\pi}\bigg)^2 \bigg(\frac{r-1}{3r}\bigg)\bigg] \\
	&\quad +2\bigg[\bigg(-\a-\frac{\hbar^2k_0^2}{8m^*}+|g|\bigg)\frac{1}{r}  +\frac{\hbar^2k_0^2}{2m^*}\bigg(\frac{K(r)}{\pi}\bigg)^2 \bigg(\frac{1+r}{3r}\bigg)\bigg]\frac{E(r)}{K(r)}, \\
	G(r) &= -\frac{\hbar^2k_0^2}{2m^*},
\end{align*}
where $E(r)$ is complete elliptic integrals of the second kind.
Here below is a convenient method to minimize the free energy density. (1) Minimize $\fed$ with respect to $\{\mu,\nu\}$ at some given $r\in(0,1)$ by solving the following equations
\begin{equation}
	\left\{
	\begin{aligned}
		\pp{\fed}{\mu} &= 4A\mu^3 + 2B\mu\nu^2 + 2D\mu + G\nu = 0, \\
		\pp{\fed}{\nu} &= 4C\nu^3 + 2B\nu\mu^2 + 2F\nu + G\mu = 0,
	\end{aligned}
	\right.
\end{equation}
and check the positive-definiteness of Hessian matrix. Denote the obtained solution as $\{\mu^*,\nu^*\}$, we compute free energy density $\fed(\mu^*,\nu^*;r)$. (2) Repeat such process at different $r$'s, and find the minimum of $\fed$ at some $r^*$. This could be realized with the help of a loop program. Then we determine the optimized order parameters $\psi_{l}(x)|_{(\mu^*,\nu^*;r^*)}$ and compute the free energy density $\fed_\m{Anal.}=\fed(\mu^*,\nu^*;r^*)$ for our analytical approximate Bloch SC solution. Finally, we compare $\fed_\m{Anal.}$ with $\fed_\m{Num.}$, the free energy density of Bloch SC solution obtained from numerical study. For various sets of parameters $(|g|/|\a|,H)$, these two results always fit well, with an error less than $0.01\%$. Part of the comparison results between the free energy density $\fed_\m{Num.}$ obtained numerically based on discrete GL equations Eq.~(9) in the main text and the analytical one $\fed_\m{Anal.}$ based on approximate trial solution for Bloch SC solution mentioned above are listed in Tab.~\ref{table}.

\begin{table}[htbp]
	\caption{Part of the comparison results $\fed_\m{Num.}$ and $\fed_\m{Anal.}$.}
	\label{table}
	\begin{tabular}{p{1.5cm}|p{1.5cm}|p{2cm}|p{2cm}|p{2cm}}
		\hline
		\hline
		\makecell[c]{$|g|/|\a|$} & 
		\makecell[c]{$\dfrac{H}{H_{\m{c}2\perp}^\m{ML}}\cdot\dfrac{a}{\xi}$} & 
		\makecell[c]{$\fed_\m{Num.}/n_0|\a|$} & 
		\makecell[c]{$\fed_\m{Anal.}/n_0|\a|$} & 
		\makecell[c]{error} \\ 
		\hline
		\makecell[c]{0.5} & \makecell[c]{0.5} & \makecell[c]{$-1.678071$} & \makecell[c]{$-1.677578$} & \makecell[c]{0.009901\%} \\
		\makecell[c]{0.5} & \makecell[c]{1.0} & \makecell[c]{$-1.319053$} & \makecell[c]{$-1.318603$} & \makecell[c]{0.009901\%} \\
		\makecell[c]{0.5} & \makecell[c]{2.0} & \makecell[c]{$-1.105187$} & \makecell[c]{$-1.105169$} & \makecell[c]{0.009901\%} \\
		\makecell[c]{0.5} & \makecell[c]{3.0} & \makecell[c]{$-1.050921$} & \makecell[c]{$-1.050920$} & \makecell[c]{0.009901\%} \\
		\makecell[c]{1.0} & \makecell[c]{0.5} & \makecell[c]{$-3.030810$} & \makecell[c]{$-3.029804$} & \makecell[c]{0.009901\%} \\
		\makecell[c]{1.0} & \makecell[c]{1.0} & \makecell[c]{$-2.189501$} & \makecell[c]{$-2.186640$} & \makecell[c]{0.009901\%} \\
		\makecell[c]{1.0} & \makecell[c]{2.0} & \makecell[c]{$-1.431888$} & \makecell[c]{$-1.431592$} & \makecell[c]{0.009901\%} \\
		\makecell[c]{1.0} & \makecell[c]{3.0} & \makecell[c]{$-1.208530$} & \makecell[c]{$-1.208515$} & \makecell[c]{0.009901\%} \\
		\makecell[c]{2.0} & \makecell[c]{0.5} & \makecell[c]{$-7.187062$} & \makecell[c]{$-7.144779$} & \makecell[c]{0.009901\%} \\
		\makecell[c]{2.0} & \makecell[c]{1.0} & \makecell[c]{$-5.314610$} & \makecell[c]{$-5.305759$} & \makecell[c]{0.009901\%} \\
		\makecell[c]{2.0} & \makecell[c]{2.0} & \makecell[c]{$-2.862650$} & \makecell[c]{$-2.859161$} & \makecell[c]{0.009901\%} \\
		\makecell[c]{2.0} & \makecell[c]{3.0} & \makecell[c]{$-1.904214$} & \makecell[c]{$-1.903928$} & \makecell[c]{0.009901\%} \\
		\makecell[c]{3.0} & \makecell[c]{0.5} & \makecell[c]{$-13.225541$} & \makecell[c]{$-13.210270$} & \makecell[c]{0.009901\%} \\
		\makecell[c]{3.0} & \makecell[c]{1.0} & \makecell[c]{$-10.208868$} & \makecell[c]{$-10.196057$} & \makecell[c]{0.009901\%} \\
		\makecell[c]{3.0} & \makecell[c]{2.0} & \makecell[c]{$-5.521574$} & \makecell[c]{$-5.508925$} & \makecell[c]{0.009901\%} \\
		\makecell[c]{3.0} & \makecell[c]{3.0} & \makecell[c]{$-3.253362$} & \makecell[c]{$-3.251645$} & \makecell[c]{0.009901\%} \\
		\hline
		\hline
	\end{tabular}
\end{table}

We could restore two special solutions at $H=0$ and $g=0$ discussed before we carry out the generic Bloch solution from this analytical expression Eq.~\eqref{anafe}. (1) As $H$ very close to 0, we expand the coefficients near $r=1$, giving $A=B=D=G=0$ and $C=\b$, $F=2(\a-|g|)$, thus the free energy density reads
\begin{equation}
	\fed = \b\nu^4 + 2(\a-|g|)\nu^2.
\end{equation}
All terms related with $\mu$ vanishes. Minimization gives $\nu^2=-\b^{-1}(\a-|g|)$, and the minimum free energy density is
\begin{equation}
	\fed|_{H=0} = -\frac{1}{\b}(\a-|g|)^2.
\end{equation}
Mathematically $\sn(ux;r)\to\tanh(ux)$ as $r\to1$ and $u\to\infty$ as $H\to0$, so $\psi_l(x)|_{H\to0}\to i\eta_{l}\e^{i\vp_l}\sqrt{-\b^{-1}(\a-|g|)}$. Hence we deduce that strictly at $H=0$, the only solution is an FF state solution, with superfluid density $\rho_{\m{s}}^{\m{FF}}|_{H=0}=-\b^{-1}(\a-|g|)$. 

(2) As $g$ very close to zero, we expand these coefficients in terms of series in $r$ near $r=0$. Up to leading order, we have
\begin{align*}
	A &= C = \frac{3}{8}\b, \quad B = \frac{\b}{4}, \\
	D &= F = \a+\frac{\hbar^2k_0^2}{4m^*}, \\
	G &= -\frac{\hbar^2k_0^2}{2m^*}.
\end{align*}
This gives exactly the free energy density at $g=0$
\begin{align}
	\fed = \frac{3}{8}\b\left(\mu^4+\nu^4\right) + \frac{\b}{4}\mu^2\nu^2 + \left(\a+\frac{\hbar^2k_0^2}{4m^*}\right)\left(\mu^2+\nu^2\right) -\frac{\hbar^2k_0^2}{2m^*}\mu\nu.
\end{align}
Minimize the free energy density by setting $\pt{\fed}/{\pt\mu}=\pt{\fed}/{\pt\nu}=0$ in this condition, yielding
\begin{equation*}
	\left\{
	\begin{aligned}
		\frac{3}{2}\b\mu^3 + \frac{\b}{2}\mu\nu^2 + \left(2\a+\frac{\hbar^2k_0^2}{2m^*}\right)\mu - \frac{\hbar^2k_0^2}{2m^*}\nu &= 0, \\
		\frac{3}{2}\b\nu^3 + \frac{\b}{2}\nu\mu^2 + \left(2\a+\frac{\hbar^2k_0^2}{2m^*}\right)\nu - \frac{\hbar^2k_0^2}{2m^*}\mu &= 0.
	\end{aligned}	
	\right.
\end{equation*}
It could be proved that real number solution for $\mu\ne\nu$ does not exist and the only non-trivial solution reads
\begin{equation*}
	\mu = \nu = \sqrt{-\frac{\a}{\b}} = \sqrt{\rho_{\m{s}0}}.
\end{equation*}
Mathematically $\sn(ux;r)\to\sin(ux)$ and $\cn(ux;r)\to\cos(ux)$ as $r\to0$. Note that $K(0)=\pi/2$, thus we have $u=2K(r)/a_0=k_0/2$ from periodic condition. Then the solution at $g=0$ reads $\psi_l(x)|_{g=0}=\sqrt{\rho_{\m{s}0}}\,\e^{i\eta_{l}\frac{k_0}{2}x}$, exactly an decoupled SC state, consistent with Eq.~(6) in the main text. And the free energy density is $\fed|_{g=0}=-\a^2/\b$. 

So far we could conclude that strictly at $H=0$, physically the system only have an FF state solution. At $g=0$, the ground state of the system is an decoupled SC state with, while the FF state solution serves as an excited state which survives up to $H_\m{c}^*(g=0)=\Phi_0/\pi a\xi$. For large enough $H$, we could assume $|g|\ll\eps_H=\hbar^2k_0^2/8m^*$ in the coefficients of Eq.~\eqref{anafe} and obtain an approximate solution at $r\to0$, very close to an decoupled SC state, as Fig.~\ref{op200} implies.

With the analytical expression of the approximate Bloch SC solution Eq.~\eqref{approx}, we could calculate local superfluid density and supercurrent density by definition Eq.~\eqref{Jsdef}.
\begin{align*}
	\rho_{\m{s}l}(x) &= |\psi_l(x)|^2 = \mu^2\cn^2(ux;r)+\nu^2\sn^2(ux;r) \\
	J_{\m{s}l}(x) &= \frac{e^*\hbar}{2m^*i}\left(\psi_l^*(x)\pt_x\psi_l(x)-\psi_l(x)\pt_x\psi_l^*(x)\right) - \frac{e^{*2}}{m^*c}|\psi_l(x)|^2A_l \\
	&= \frac{e^*\hbar}{2m^*i}\left(2i\eta_{l}\frac{K(r)}{\pi}\frac{e^*Ha}{\hbar c}\mu\nu\dn(ux;r)\right) - \eta_{l}\frac{e^{*2}Ha}{2m^*c}\left(\mu^2\cn^2(ux;r)+\nu^2\sn^2(ux;r)\right) \\
	&= \eta_{l}\frac{e^{*2}Ha}{2m^*c}\left[\frac{2K(r)}{\pi}\mu\nu\dn(ux;r) - \left(\mu^2\cn^2(ux;r)+\nu^2\sn^2(ux;r)\right)\right].
\end{align*}
where $u=2K(r)/a_0=\frac{K(r)}{\pi}\frac{e^*Ha}{\hbar c}$. The contribution from net orbital magnetization reads
\begin{align*}
	\frac{m^{*}c}{e^{*2}}\oint\frac{\bf{J}_{\m{s}}(x)}{\rho_{\m{s}}(x)}\cdot\d\bf{s}
	&= \frac{m^{*}c}{e^{*2}}\left[\int_{0}^{a_0} \frac{J_{\m{s}1}(x)}{\rho_{\m{s}1}(x)} \d x + \int_{a_0}^{0} \frac{J_{\m{s}2}(x)}{\rho_{\m{s}2}(x)} \d x\right] \\
	&= \frac{m^{*}c}{e^{*2}}\cdot 2\int_{0}^{a_0}\frac{e^{*2}Ha}{2m^*c} \left[\frac{2K(r)}{\pi}\frac{\mu\nu\dn(ux;r)}{\mu^2\cn^2(ux;r)+\nu^2\sn^2(ux;r)} - 1\right]\d x \\
	&= Ha \int_{0}^{a_0}\left[\frac{2K(r)}{\pi}\frac{\mu\nu\dn(ux;r)}{\mu^2\cn^2(ux;r)+\nu^2\sn^2(ux;r)} - 1\right]\d x.
\end{align*}
Substituting our solution $\{\mu^*,\nu^*,r^*\}$ at any given parameters $(|g|/|\a|,H)$ into the above expression all yields net orbital magnetization of Bloch SC solution equals 0, which is in good agreement with our deduction based on numerical solution Eq.~\eqref{netmagzero}. Hence the fluxoid in one period $a_0$ is
\begin{equation}
	\Phi' = \oint\bf{A}\cdot\d\bf{s} + \frac{m^{*}c}{e^{*2}}\oint\frac{\bf{J}_{\m{s}}(x)}{\rho_{\m{s}}(x)}\cdot\d\bf{s} = \Phi_0.
\end{equation}

\subsection{Estimation of the scale of physical parameters}
We do not consider temperature dependence in our model, for any given temperature $T<T_\m{c}$, $\a(T)<0$ and is treated as a constant, whose absolute value $|\a(T)|$ serves as energy unit in our numerical solution to measure $g$. All the above theoretical derivations are expressed in Gaussian units $B=H$ for convenience, whereas in SI units, $B=\mu_0H$. Experimentally, $H_{\m{c}2\perp}^\m{ML}$ usually scales at about $5000\m{A/m}$, so $B_{\m{c}2\perp}^\m{ML}\sim6\times10^{-3}\m{T}=6\m{mT}$. 
This gives a coherence length $\xi\sim2\times10^{-7}\m{m}=200\m{nm}$ and $|\a|=\hbar^2(2m^*\xi^2)^{-1}\sim10^{-7}\m{eV}\sim10^{-4}\m{meV}$. 
If the applied in-plane magnetic field $B_{\parallel}$ scales as $B_{\m{c}2\perp}^\m{ML}$, about $1\sim10\m{mT}$, and the interlayer spacing $a$ scales at atomic level about $10\sim100\AA$ or $1\sim10\m{nm}$, the period of local superfluid density $\rho_{\m{s}l}(x)=|\psi_{l}(x)|^2$, namely $a_0=\Phi_0/B_{\parallel}a$, would scale about $\sim10^{-6}\m{m}=1\mu\m{m}$. Besides, and Josephson coupling energy $g$ is measured in unit $|\a|$ and sscales about $10^{-4}\m{meV}$.

\section{Stability of the Bilayer Superconducting System in the Presence of a Tilted Magnetic Field}
\subsection{General solution of critical perpendicular magnetic field}
As mentioned in the main text, in our numerical study on the bilayer SC system in the presence of only an in-plane magnetic field, mathematically the Bloch solution exists throughout the whole phase diagram. Therefore an actual parallel critical field $H_{\m{c}\parallel}$ will be determined physically by the parallel critical field for a monolayer thin film $H_{\m{c}\parallel}^\m{ML}=2\sqrt{3}H_{\m{c}2\perp}^\m{ML}\cdot(\xi/d)$~\cite{Tinkham} and the Pauli paramagnetic limit $H_{\m{P}}$~\cite{Chandrasekhar,Clogston}, where $d$ is the thickness of each thin film. The Bloch SC state will vanish when $H>H_{\m{c}\parallel}$. 

Here we could raise a new issue: What is the stability of this bilayer SC system in the presence of a tilted magnetic field? The term ``tilted" means the magnetic field has two components $\Hpara$ and $\Hperp$, parallel and perpendicular to the bilayer thin films, respectively. If $\Hperp=0$, the problem restores to our main study. Whereas if $\Hpara=0$, the Josephson tunneling effect comes from the weak link between two SC ultra-thin films might influence the perpendicular critical field $H_{\m{c}2\perp}^\m{BL}$.

Consider applying an extra $\Hperp$ to the system consisting of two layers of superconducting ultra-thin films in the presence of an pre-applied parallel magnetic field, we find a general theory on the perpendicular critical field $H_{\m{c}2\perp}^\m{BL}$ at given parameters $(|g|/|\a|,\Hpara)$, which tells us the stability of this bilayer system, either in generic Bloch SC state or in FF state originally, under this additional component $\Hperp$. The order parameters might have some other exotic spatial modulation but we only focus on the critical point, beyond which the SC state is killed by $\Hperp$. The total magnetic field could be expressed as $\bf{H} = \Hpara\ey + \Hperp\ez$, again we choose Landau gauge to express the vector potential as $\bf{A}=\Hpara z\ex+\Hperp x\ey$. The vector potential on each layer reads
\begin{equation}
	\bf{A}_l = \eta_{l}\Hpara\frac{a}{2}\ex + \Hperp x\ey,
\end{equation}
where $a$ is the distance between two layers. We could always choose the direction of $\Hpara$ and $\Hperp$ as $\ey$ and $\ez$, so their values are set to be positive below. Near $H_{\m{c}2\perp}^\m{BL}$ the superfluid density $|\psi_{l}|^2$ is small enough, and the nonlinear $\b$-term vanishes in Eq.~(3) in the main text, thus we could directly start from the linearized GL equations
\begin{equation}
	\a\psi_l(x,y) + \frac{1}{2m^*}\left(\frac{\hbar}{i}\nabla-\frac{e^*}{c}\bf{A}_l\right)^2\psi_l(x,y) + g\psi_{\bar{l}}(x,y) = 0,
\end{equation}
Define $\hat{H}_l=\bm{\Pi}_l^2/2m^*$, where the canonical momentum is
\begin{equation}
	\bm{\Pi}_l = \frac{\hbar}{i}\nabla-\frac{e^*}{c}\bf{A}_l. \quad (l=1,2)
\end{equation}
The linearized GL equations could be expressed in the following matrix form
\begin{equation}\label{glmat}
	\begin{pmatrix}
		\hat{H}_1 & g \\
		g & \hat{H}_2
	\end{pmatrix}
	\begin{pmatrix}
		\psi_1 \\
		\psi_2
	\end{pmatrix}
	= -\a
	\begin{pmatrix}
		\psi_1 \\
		\psi_2
	\end{pmatrix}.
\end{equation}
This matrix operator equation could be regarded as a ``Schr\"{o}dinger" equation $\hat{H}\Psi=E\Psi$, where $\Psi=(\psi_1,\psi_2)^{\m{T}}$ and the ``Hamiltonian" matrix takes the form
\begin{equation}
	\hat{H} = 
	\begin{pmatrix}
		\hat{H}_1 & g \\
		g & \hat{H}_2
	\end{pmatrix},
\end{equation}
However, our treatment is somewhat different from solving an ordinary quantum mechanical equation problem. Now the ``Hamiltonian" $\hat{H}$ is dependent on parameters $\bf{H}$ and $g$, whose eigenvalue $E$ is fixed to $-\a$. Our target is to obtain maximum $\Hperp$ at given parameters $\Hpara$ and $g$. Note that $\hat{p}_y=-i\hbar\pt_{y}$ commutes with the Hamiltonian. Then
\begin{align}\label{shoham}
	\hat{H}_{l} &= \frac{1}{2m^*}\left(\frac{\hbar}{i}\pp{}{x}-\eta_{l}\frac{e^*\Hpara a}{2c}\right)^2 + \frac{1}{2m^*}\left(\frac{\hbar}{i}\pp{}{y}-\frac{e^*\Hperp}{c}x\right)^2 \notag\\
	&= \frac{1}{2m^*}\left(\hat{p}_x - \eta_{l}\frac{\hbar k_0}{2}\right)^2 + \frac{1}{2}m^*\wperp^2\left(\hat{x}-x_0\right)^2,
\end{align} 
where we define the ``cyclotron frequencies" $\om_{\m{c}\parallel(\perp)}=e^*H_{\m{c}\parallel(\perp)}/m^*c$, and ``center shift" $x_0=p_yc/e^*\Hperp$. Here $k_0=2\pi\Hpara a/\Phi_0$ is so defined in agreement with our main study.
Define $\hat{\z}\equiv\hat{x}-x_0$ and $\hat{p}_\z=-i\hbar\pt_\z=-i\hbar\pt_x=\hat{p}_x$, we have
\begin{equation*}
	\hat{H}_{l} = \frac{\hat{p}_\z^2}{2m^*} + \frac{1}{2}m^*\wperp^2\hat{\z}^2 - \eta_{l} \frac{\hbar k_0}{2m^*}\hat{p}_\z + \frac{\hbar^2k_0^2}{8m^*}.
\end{equation*} 
This is exactly a Hamiltonian operator of a dissipative harmonic oscillator, thus we could re-express the ``Hamiltonian" in terms of bosonic ladder operators
\begin{equation}\label{ladder}
	\hat{H} = \om\left(\had\ha+\frac{1}{2}\right)\tau_0 - v(\had+\ha)\tau_3 + g\tau_1 + \eps_H\tau_0,
\end{equation}
where $\tau_1,\tau_3$ are Pauli matrices for layer degree of freedom and $\tau_0$ is identity matrix. The coefficients are defined as $\om \equiv \hbar\wperp$, $\eps_H \equiv \hbar^2k_0^2/8m^* = m^*\wpara^2a^2/8$ and $v=\sqrt{\eps_H\om}$ is related with both parallel and perpendicular magnetic fields. To show the detail, the ladder operators are defined as $\ha=\e^{-i\pi/2}\hb$, $\had=\e^{i\pi/2}\hbd$, where $\hb=(2m^*\hbar\wperp)^{-1/2}(m^*\wperp\hat{\z}+i\hat{p}_\z)$. The eigenvalues of $\hat{H}$ are unchanged under a unitary transformation $U=\exp(i\pi\tau_2/4)$, and we move the constants to the right hand side of our original ``Schr\"{o}dinger" equation, yielding $\hat{H}_\m{R}\Psi_\m{R}=E_\m{R}\Psi_\m{R}$, where
\begin{equation}\label{Rabiham}
	\hat{H}_\m{R} = \om\had\ha\tau_0 + v(\had+\ha)\tau_1 + g\tau_3.
\end{equation}
This is exactly the so-called Rabi model \cite{Rabi}, which is often discussed in quantum optics. Rabi model describes a harmonic oscillator (bosonic mode) coupled to a ``spin" or two-level system. Here $v$ seems to be ``coupling strength" between them, and $g$ is the ``level splitting", which is actually the interlayer Josephson tunneling in our realistic model. 

In our study, we first solve the Rabi model at given parameters $(g,\Hpara)$ and various $\Hperp$, obtaining the eigenvalues $E_{\m{R}}(\om)$, which is related with $\Hperp$, then we solve the following equation
\begin{equation}\label{solvew}
	E_\m{R}(\om) = -\left(\a+\frac{\om}{2}+\eps_H\right)
\end{equation}
for all eigenvalues and obtain a set of values $\om$. The allowed maximum value of $\om=\hbar\wperp$ hence gives $H_{\m{c}2\perp}$. However, Rabi model does not have explicit analytical solutions, so we have to solve it numerically \cite{RabiSol,JCsol}. In this research, we apply exact diagonalization to the Rabi model with a truncation up to $N_\m{trunc}=50$ in occupation number of the bosonic mode. Then we solve $\om$ with Eq.~\eqref{solvew} for all eigenvalues, and the maximum $\om$ among all the solutions is expected. Practically, using the first few eigenvalues is enough to obtain $\om_{\max}$, and hence $H_{\m{c}2\perp}^\m{BL}$. We measure coupling energy $g$ in units $|\a|$, and parallel magnetic field in dimensionless value $\gamma=(\Hpara/H_{\m{c}2\perp}^\m{ML})\cdot(a/\xi)$. The perpendicular magnetic field could be measured directly in unit $H_{\m{c}2\perp}^\m{ML}$. We study how perpendicular critical field of this weak-linked bilayer SC system, $H_{\m{c}2\perp}^\m{BL}$, changes with in-plane magnetic field $\Hpara$ and Josephson tunneling $g$. The results are shown in Fig.~\ref{hperp}.

\begin{figure}[htbp]
	\centering
	\subfigure{
		\label{hvsh}
		\begin{overpic}[width=0.48\linewidth]{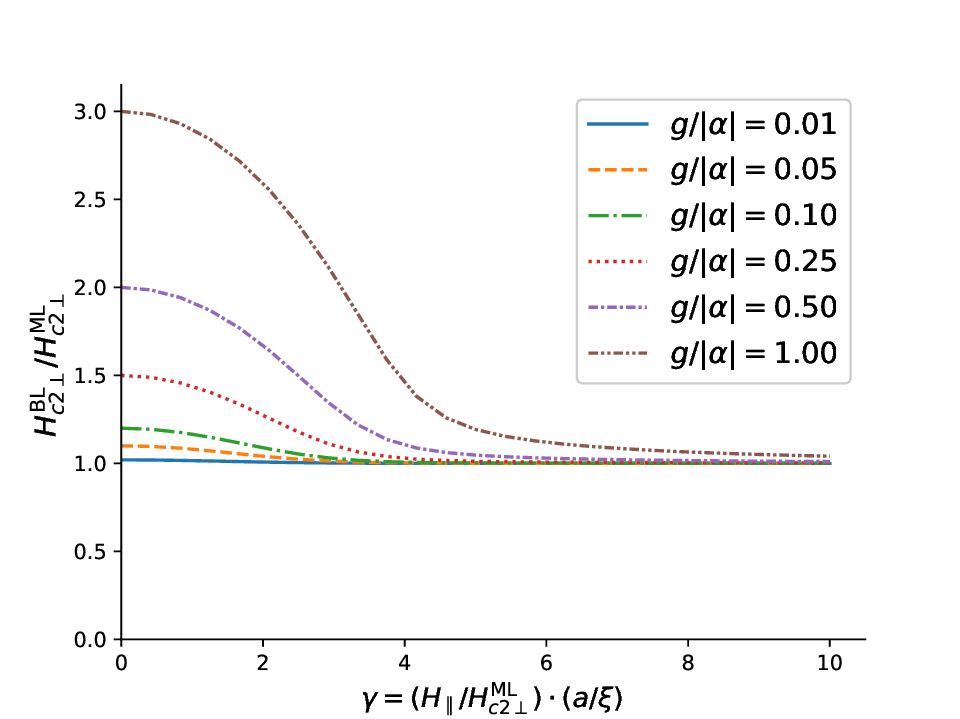}
			\put(0,65){(a)}
		\end{overpic}
	}
	\hfill
	\subfigure{
		\label{hvsg}
		\begin{overpic}[width=0.48\linewidth]{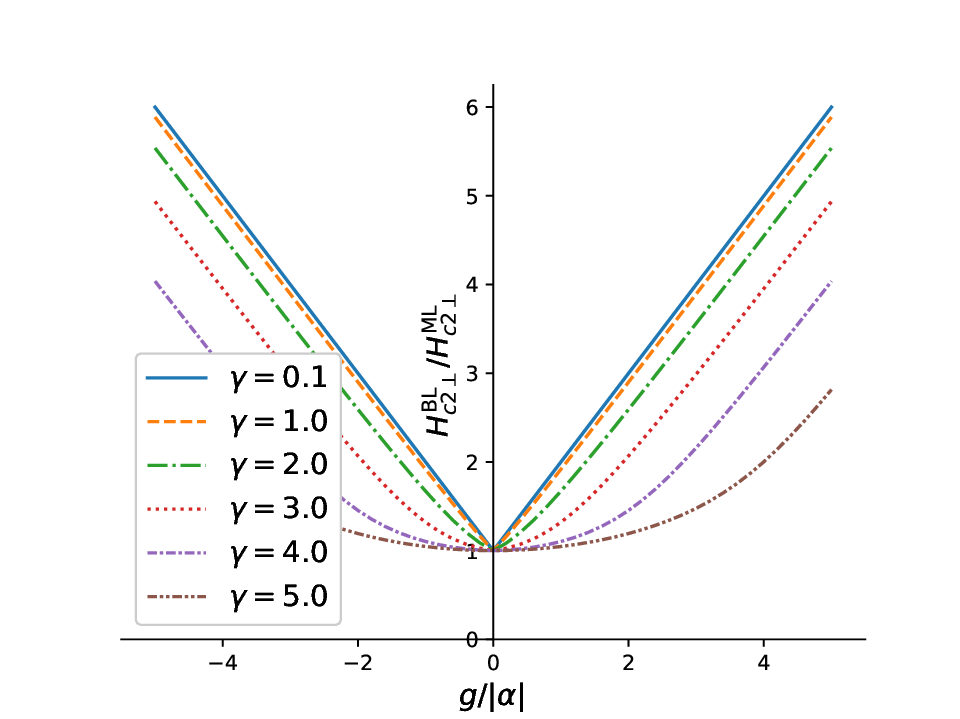}
			\put(5,65){(b)}
		\end{overpic}
	}
	\caption{Numerical results for perpendicular upper critical field, as functions of $\Hpara$ and $g$. (a) Plot of $H_{\m{c}2\perp}^\m{BL}$ as a function of $\Hpara$. (b) Plot of $H_{\m{c}2\perp}^\m{BL}$ as a function of $g$, where the legends represent dimensionless value $\g=(\Hpara/H_{\m{c}2\perp}^{\mathrm{ML}})\cdot(a/\xi)$.}
	\label{hperp}
\end{figure}

\subsection{Analytical approach at specific limits}

(1) Large-$g$ limit, for $|g|\gg\eps_H$. Assume $|g|$ is comparable with $|\a|,\om$ and far greater than $v$, which means $\eps_H\ll\om,|g|$, rotation wave approximation (RWA) could be carried out and we are led to an analytically solvable Jaynes-Cummings (JC) model \cite{JCmodel}
\begin{equation}\label{jc}
	\hat{H}_\m{JC} = \om\had\ha + v(\had\tau^-+\ha\tau^+) + g\tau_3,
\end{equation}
where $\tau^{\pm}=(\tau_1\pm i\tau_2)/2$. From $\eps_H\ll|\a|,\om$, we derive the limiting condition for the applied fields $\gamma=(\Hpara/H_{\m{c}2\perp}^\m{ML})\cdot(a/\xi)\ll 2$, and $\lperp \sim \mathcal{O}(\xi) \ll \lpara^2/a$, where $\ell$ is the ``cyclotron radius" and defined as $\ell_{\perp(\parallel)}=\sqrt{\hbar/m^*\om_{\m{c}\perp(\parallel)}}= \sqrt{\Phi_0/2\pi H_{\perp(\parallel)}}$. The JC model Eq.~\eqref{jc} is exactly solved and analytically we arrive at $\om_\m{max} = 2(|\a|+|g|-\eps_H)$ under this condition. Hence the perpendicular critical field reads
\begin{equation}\label{hcb1}
	H_{\m{c}2\perp}^\m{BL} = H_{\m{c}2\perp}^\m{ML}\left(1+\frac{|g|}{|\a|}\right) - \frac{2\pi}{\Phi_0}\left(\frac{\Hpara a}{2}\right)^2.
\end{equation}
Note that if $\Hpara=0$, without parallel field, for such a bilayer superconducting system, the perpendicular critical field shifts to
\begin{equation*}
	H_{\m{c}2\perp}^\m{BL} = H_{\m{c}2\perp}^\m{ML}\left(1+\frac{|g|}{|\a|}\right)
\end{equation*}
due to the interlayer Josephson tunneling $g$, compared with $H_{\m{c}2\perp}^\m{ML}$. When applying to the system a very small parallel field, the perpendicular critical field slightly decreases by last term in Eq.~\eqref{hcb1}.

(2) Small-$g$ limit, with $|g|\ll\mathcal{O}(|\a|)$. Return to our original model Eq.~\eqref{ladder}, we could rewrite it as
\begin{equation}
	\hat{H} = \hat{H}_0 + \hat{H}' + \frac{\om}{2} + \eps_H,
\end{equation}
where $\hat{H}_0 = \om\had\ha - v(\had+\ha)\s^z$ and $\hat{H}' = g\s^x$. If $g\ll\om\sim\mathcal{O}(|\a|)$, this model could be solved with the help of perturbation theory, leaving a transcendental equation $\om = 2|\a| \mp 2g\exp\left(-2\eps_H/\om\right)$ for our study. Now that $|g|\ll\om$ and $\om$ scales as $2|\a|$, the larger solution, which determines $H_{\m{c}2\perp}^\m{BL}$, should come from the equation $\om = 2|\a| + 2|g|\exp\left(-2\eps_H/\om\right)$, regardless of the sign of $g$.
For $2\eps_H\gg\om\sim2|\a|$, i.e. $\g\ll2$, we arrive at one approximate equation $\om_\m{max}=2|\a|$. For $2\eps_H\ll\om\sim2|\a|$, i.e. $\g\gg2$, we arrive at an approximate solution to this transcendental equation
\begin{equation*}
	\om_\m{max} = 2|\a| + 2|g|\exp\left(-\frac{\eps_H}{|\a|+|g|}\right).
\end{equation*}
That is, under the condition $|g|/|\a|\ll1$, if $\g \ll 2$,
\begin{subequations}\label{hcb2}
	\begin{equation}
		H_{\m{c}2\perp}^\m{BL} = \frac{\Phi_0}{2\pi\xi^2} = H_{\m{c}2\perp}^\m{ML},
	\end{equation}
	which means the Josephson tunneling and parallel magnetic field are weak enough, the bilayer system is almost two individual monolayers in the presence of a perpendicular field. Whereas if $\g \gg 2$
	\begin{equation}
		H_{\m{c}2\perp}^\m{BL} = H_{\m{c}2\perp}^\m{ML}\left\{1 + \frac{|g|}{|\a|} \exp\left[-\frac{1}{1+|g|/|\a|}\left(\frac{\pi a\xi}{\Phi_0}\Hpara\right)^2\right]\right\}.
	\end{equation}
\end{subequations}

Both Eqs.~\eqref{hcb1} and \eqref{hcb2} are in good agreement with numerical results Fig.~\ref{hperp} at corresponding limiting conditions.

\end{widetext}

\end{document}